\documentclass[aps,pre,twocolumn,floatfix,showpacs]{revtex4}
\usepackage{epsf,amsmath,amssymb,verbatim,color}
\definecolor{orange}{rgb}{1.0,0.7,0.0}
\begin{document}
\newcommand{\be}{\begin{equation}}
\newcommand{\ee}{\end{equation}}
\title{Fractality in complex networks: critical and supercritical skeletons}
\author{J.~S. Kim,$^1$ K.-I. Goh,$^2$ G. Salvi,$^1$ E. Oh,$^1$
B. Kahng,$^{1,3}$ and D. Kim$^1$} \affiliation{{$^1$School of
Physics and Astronomy and Center for Theoretical Physics, Seoul
National University, Seoul 151-747, Korea}\\
{$^2${Center for Cancer Systems Biology, Dana-Farber Cancer
Institute, Harvard Medical School, Boston, MA 02115 and}\\
{Center for Complex Network Research and Department of Physics,
University of Notre Dame, Notre Dame, IN 46556}}\\
{$^3$Center for Nonlinear Studies, Los Alamos National
Laboratory, Los Alamos, NM 87545}}
\date{\today}

\begin{abstract}
Fractal scaling---a power-law behavior of the number of boxes
needed to tile a given network with respect to the lateral size of
the box---is studied. We introduce a new box-covering algorithm
that is a modified version of the original algorithm introduced by
Song {\em et al.} [Nature (London) {\bf 433,} 392 (2005)]; this
algorithm enables effective computation and easy implementation.
Fractal networks are viewed as comprising a skeleton and
shortcuts. The skeleton, embedded underneath the original network,
is a special type of spanning tree based on the edge betweenness
centrality; it provides a scaffold for the fractality of the
network. When the skeleton is regarded as a branching tree, it
exhibits a plateau in the mean branching number as a function of
the distance from a root.  For non-fractal networks, on the other
hand, the mean branching number decays to zero without forming a
plateau. Based on these observations, we construct a fractal
network model by combining a random branching tree and local
shortcuts. The scaffold branching tree can be either critical or
supercritical, depending on the small-worldness of a given
network. For the network constructed from the critical
(supercritical) branching tree, the average number of vertices
within a given box grows with the lateral size of the box
according to a power-law (an exponential) form in the
cluster-growing method. The critical and supercritical skeletons
are observed in protein interaction networks and the world-wide
web, respectively. The distribution of box masses, i.e., the
number of vertices within each box, follows a power law
$P_m(M)\sim M^{-\eta}$. The exponent $\eta$ depends on the box
lateral size $\ell_B$. For small values of $\ell_B$, $\eta$ is
equal to the degree exponent $\gamma$ of a given scale-free
network, whereas $\eta$ approaches the exponent
$\tau=\gamma/(\gamma-1)$ as $\ell_B$ increases, which is the
exponent of the cluster-size distribution of the random branching
tree. Finally, we study the perimeter $H_{\alpha}$ of a given box
$\alpha$, i.e., the number of edges connected to different boxes
from a given box $\alpha$ as a function of the box mass
$M_{B,\alpha}$. It is obtained that the average perimeter over the
boxes with box mass $M_B$ is likely to scale as $\langle H (M_B)
\rangle \sim M_B$, irrespective of the box size $\ell_B$.
\end{abstract}
\maketitle

\section{Introduction}
Fractal scaling recently observed~\cite{ss} in real-world
scale-free (SF) networks such as the world-wide web (WWW)
\cite{www}, metabolic network of {\em Escherichia coli} and other
microorganisms \cite{metabolic}, and protein interaction network
of {\it Homo sapiens}~\cite{dip} has opened a new perspective in
the study of networks. SF networks~\cite{ba} are those that
exhibit a power-law degree distribution $P_d(k)\sim k^{-\gamma}$.
Degree $k$ is the number of edges connected to a given vertex.
Fractal scaling implies a power-law relationship between the
minimum number of boxes $N_B(\ell_B)$ needed to tile the entire
network and the lateral size of the boxes $\ell_B$, i.e., \be
N_B(\ell_B)\sim \ell_B^{-d_B}, \label{fractal} \ee where $d_B$ is
the fractal dimension~\cite{feder}. This power-law scaling implies
that the average number of vertices $\langle M_B(\ell_B) \rangle$
within a box of lateral box size $\ell_B$ scales according to a
power law as \be \langle M_B(\ell_B)\rangle \sim \ell_B^{d_B}.
\label{boxcovering}\ee Here, the relation of system size $N \sim
N_B(\ell_B)\langle M_B(\ell_B) \rangle$ is used. This counting
method is called the box-covering method. At a glance, the
power-law fractal scaling (\ref{fractal}) is not consistent with
the notion of small-worldness (SW) of SF networks. SW implies that
the average number of vertices within a distance $\ell_C$ from a
vertex scales as \be \langle M_C(\ell_C) \rangle \sim
e^{\ell_C/\ell_0}, \label{sw} \ee where $\ell_0$ is a constant.
This counting method is called the cluster-growing method. Here,
subscripts $B$ and $C$ represent the box-covering and
cluster-growing methods, respectively. The number of vertices $M$
within a box is referred to as the box mass. This contradiction
can be resolved by the fact that a vertex is (can be) counted only
once (more than once) in the box-covering method (in the
cluster-growing method).

Recently, it was suggested that the fractal scaling originates
from the disassortative correlation between two neighboring
degrees \cite{yook} or the repulsion between hubs \cite{song2}.
Moreover, we showed~\cite{goh2006} that the fractal network
contains the fractal skeleton \cite{skeleton} underneath it; this
skeleton is a special type of spanning tree, formed by edges with
the highest betweenness centralities \cite{freeman,gn} or
loads~\cite{static}. The remaining edges in the network are
referred to as shortcuts that contribute to loop formation. The
skeleton of an SF network also follows a power-law degree
distribution, where its degree exponent can differ slightly from
that of the original network~\cite{skeleton}. For fractal networks
that follow fractal scaling (\ref{fractal}), we have shown
\cite{goh2006} that each of their skeletons exhibits fractal
scaling similar to that of the original network. The number of
boxes needed to cover the original network is almost identical to
that needed to cover the skeleton. Thus, since the skeleton is a
simple tree structure, it is more useful than the original network
for studying the origin of the fractality.

It was shown~\cite{goh2006} that the skeleton of the fractal
network exhibits a non-dying branching structure, referred to as a
{\em persistent} branching structure hereafter. A skeleton can be
considered as a tree generated in a branching process
\cite{harris} starting from the root vertex. This mapping can be
applied to any tree. If a branching process occurs in an
uncorrelated manner, the branching tree obtained from it exhibits
a plateau, albeit fluctuating, in the mean branching number
function $\bar{n}(d)$, which is defined as the average number of
offsprings created by vertices at a distance $d$ from the root.
Actually, the plateau is formed when $\bar{n}(d)$ is independent
of $d$; this is denoted as $\bar n$ for future discussions. The
branching tree structure obtained from the random branching
process is known to be a fractal for the critical case
\cite{burda,dslee}, where the mean branching rate is $\langle
n\rangle=1$. Here, $\langle n \rangle$ is defined as \be \langle
n\rangle\equiv\sum_{n=0}^{\infty}nb_n, \label{mbr}\ee where $b_n$
is the probability that a vertex will produce $n$ offsprings in
each step. Thus, $\langle n \rangle={\bar n}$ for the random
branching tree. The fractal dimension of the SF branching tree
generated with branching probability $b_n\sim n^{-\gamma}$ is
given by \cite{burda,dslee} \be d_B=
\begin{cases}
 ({\gamma-1})/({\gamma-2}) & \text{for}~2 < \gamma < 3,\\
 2 & \text{for}~\gamma > 3.
\end{cases}
\label{z} \ee Thus, the presence of the skeleton as the critical
branching tree served as a scaffold for the fractality of the
fractal network. Then, the original fractal network is a dressed
structure to the skeleton with local shortcuts; the number of
shortcuts is kept minimal in order to ensure fractality. This idea
is demonstrated by observing that the number of boxes in the
fractal scaling (\ref{fractal}) for an original fractal network is
similar to that of its skeleton \cite{goh2006}. When $\langle n
\rangle >1$, we will show that a supercritical branching tree is
also a fractal from the perspective of the fractal scaling
(\ref{fractal}), although exponential relation in the average box
mass (\ref{sw}) holds in the cluster-growing method. The
supercritical branching tree also exhibits a plateau in
$\bar{n}(d)$. Based on these observations, we define a network to
be a fractal if (i) it exhibits a power-law scaling Eq.~(1) in the
box-covering method {\em and} (ii) its skeleton is also a fractal
with the persistent branching structure, i.e., a plateau exists in
the mean branching number function $\bar{n}(d)$.

Based on these findings, we introduced a fractal network model by
incorporating the random critical branching tree and local
shortcuts \cite{goh2006}. In this paper, we will show that the
model can also be generalized for the supercritical branching
tree, thereby facilitating a better understanding of fractal
networks. For example, the model based on the supercritical
branching tree can explain the subtle coexistence of SW and
fractality as observed in the WWW.

Due to the heterogeneity of the degree distribution in the SF
networks, the distribution of box masses is also nontrivial. It
exhibits a power-law tail with exponent $\eta$, \be P_m(M_B)\sim
M_B^{-\eta}, \label{PmMB0} \ee in the box-covering method, whereas
it exhibits a peak at a characteristic mass in the cluster-growing
method. In this paper, we perform a detailed analysis of the
real-world fractal network as well as the fractal network model,
thereby showing that the box-mass distribution in the box-covering
method for the fractal network can be explained by the branching
dynamics. The exponent $\eta$ of the box-mass distribution is
related to the exponent $\tau$ that describes the size
distribution of random branching trees \cite{harris}. In
particular, for the critical SF branching tree, $\tau$ is known to
be \cite{dslee,saichev} \be \tau=
\begin{cases}
 {\gamma}/({\gamma-1}) & \text{for}~2 < \gamma < 3,\\
 3/2 & \text{for}~\gamma > 3.
\end{cases}
\label{tau} \ee The same value of $\tau$ can be derived for the
supercritical SF branching tree; however, the power-law scaling
behavior is limited to a finite characteristic size depending on
$\langle n\rangle$ and $\gamma$ \cite{dslee2006}. Thus, the
cluster-size distribution follows a power law for both the
critical branching tree and the supercritical branching tree up to
the characteristic size.

In the first part of this paper, we present the fractal property
of real-world networks and the model of the fractal network in
detail as well as a further analysis of our previous work
\cite{goh2006}. Initially, in Sec.~II we introduce a modified
version of the box-covering method employed in this paper,
following which we present the fractal scaling (Sec.~III) and the
mean branching number analysis (Sec.~IV) for an extended list of
complex networks, real-world and model networks. In the later
part, we provide a general description of a model of fractal SF
networks including the supercritical branching tree and study its
property in detail in Sec.~V. In Sec.~VI, we examine the average
box mass and the box-mass distribution for the fractal networks in
the box-covering and cluster-growing methods. The average
perimeter of a box as a function of the box mass is studied in
Sec.~VII. The summary follows in Sec.~VIII.

\section{The box-covering method}
The fundamental relation of fractal scaling (\ref{fractal}) is
based on the procedure referred to as the box-covering method
\cite{ss} that calculates the number of boxes $N_B$ needed to
cover the entire network with boxes of lateral size $\ell_B$. This
is analogous to the box-counting method normally used in fractal
geometry \cite{feder}. Song {\it et al.} \cite{ss} introduced a
new definition of the box applicable to complex networks such that
the maximum separation between any pair of vertices within each
box is less than $\ell_S$. However, this particular definition has
proved to be inessential for fractal scaling. Throughout this
study, we utilize a different version of the box-covering method
introduced here; this method involves sequential steps of box
covering, thereby providing an easy implementation:
\begin{enumerate}
\item[(i)] Select a vertex randomly at each step; this vertex serves
as a seed.
\item[(ii)] Search the network by distance $\ell_B$ from the seed
and assign {\em newly burned vertices} to the new box. If no new
vertex is found, do nothing.
\item[(iii)] Repeat {(i)} and {(ii)} until all vertices
are assigned to their respective boxes.
\end{enumerate}
\begin{figure}[t]
\centerline{\epsfxsize=5cm \epsfbox{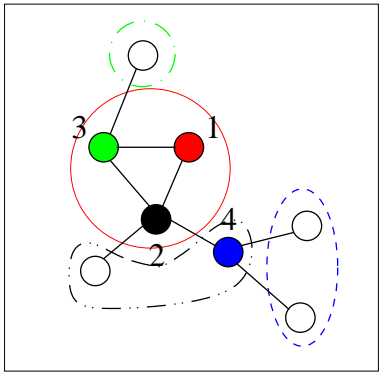}} \caption{ (Color
online) Schematic illustration of the box-covering algorithm
introduced in this work. Vertices are selected randomly, for
example, from vertex 1 to 4 successively. Vertices within distance
$\ell_B=1$ from vertex 1 are assigned to a box represented by the
solid (red) circle. Vertices from vertex 2, not yet assigned to
their respective box are represented by the dash-dot-dot (black)
closed curve, vertices from vertex 3 are represented by dash-dot
(green) circle and vertices from vertex 4 are represented by the
dashed (blue) ellipse.} \label{box_method}
\end{figure}
\begin{figure}[t]
\centerline{\epsfxsize=8cm \epsfbox{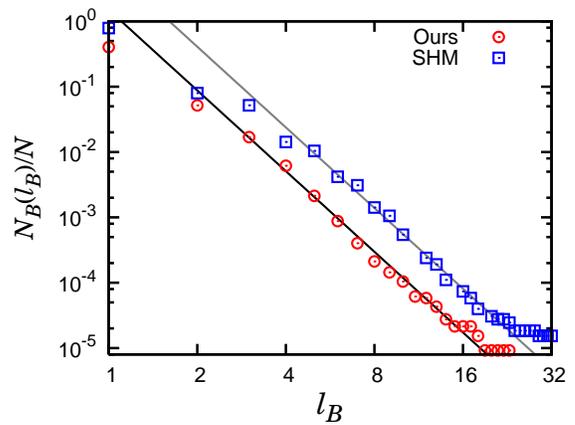}} \caption{ (Color
online) Comparison of the box-covering methods introduced by Song
{\it et al.}~\cite{ss} (\textcolor{blue}{$\square$}) and in this
paper (\textcolor{red}{$\circ$}). The results obtained from the
two box-covering methods applied to the WWW are plotted here. The
two methods yield the same fractal dimension $d_B \approx$ 4.1.}
\label{frac}
\end{figure}
\begin{table*}
\caption{Properties of real-world networks studied in this work.
For each network, the number of vertices $N$, the average degree
$\langle k\rangle$, the assortativity mixing index $r$, the
average separation $\langle d\rangle$ of all pairs of vertices,
and the maximum separation $d_{\rm max}$ among all pairs of
vertices are tabulated.}
\begin{ruledtabular}
\begin{tabular}{lllllll}
Name & $N$ & $\langle k\rangle$ & $r$ & $\langle d\rangle$ &
$d_{\rm max}$ & category
\\
\hline World-wide web &
325729 & 6.7 & $-0.05$ & 7.2 & 46 & fractal and SW\\
Metabolic network of {\it E. coli} &
2859 & 4.8 & $-0.16$ & 4.7 & 18 & fractal and SW\\
PIN of {\it H. sapiens} &
563 & 3.1 & $-0.14$ & 6.9 & 21 & fractal but not SW\\
PIN of {\it S. cerevisiae} &
741 & 4.7 & $\phantom{-}0.41$ & 10.8 & 27 & fractal but not SW\\
Actor network &
374511 & 80.2 & $\phantom{-}0.22$ & 3.7 & --- & non-fractal and SW\\
Coauthorship network (cond-mat) &
13861 & 6.4 & $\phantom{-}0.16$ & 6.6 & 18 & non-fractal and SW\\
Internet at the AS level &
16644 & 4.3 & $-0.20$ & 3.7 & 10 & non-fractal and SW\\
Router network &
284805 & 3.2 & $-0.01$ & 8.8 & 30 & non-fractal and SW \\
Power grid of the USA &
4941 & 4.9 & $\phantom{-}0.06$ & 8.5 & 17 & undetermined and SW \\
\end{tabular}
\end{ruledtabular}
\end{table*}
The above method is schematically illustrated in
Fig.~\ref{box_method}. It should be noted that vertices can be
disconnected within a box, but connected through a vertex (or
vertices) in a different box (or boxes) as in the case of box 2 as
shown in Fig.~\ref{box_method}. On the other hand, if we construct
a box with only connected vertices, the power-law behavior
Eq.~(\ref{fractal}) is not observed. The box size $\ell_B$ used
here is related to $\ell_S$ approximately as $\ell_S \approx
2\ell_B+1$. A different Monte Carlo realization of this procedure
((i)--(iii)) yields a different number of boxes for covering the
network. In this study, for simplicity, we choose the smallest
number of boxes among all the trials. Although, this algorithm
provides equivalent fractal dimension $d_B$ to the one introduced
by Song {\it et al.} \cite{ss}, it is easier to implement and is
effective in computation. In Fig.~\ref{frac}, we compare the two
box-covering methods applied to the WWW, demonstrating that the
same fractal dimension $d_B$ is obtained. It should be noted that
the box number $N_B$ we employ is not the {\em minimum} number
among all the possible tiling configurations. Finding the actual
minimum number over all configurations is a challenging task.
However, in this paper, we focus on the problem of the fractal
scaling within the framework of the box-covering algorithm
introduced above.

\section{Fractal scaling analysis}

\begin{figure*}
\begin{minipage}{.5\linewidth}
\centerline{\epsfxsize=9.5cm \epsfbox{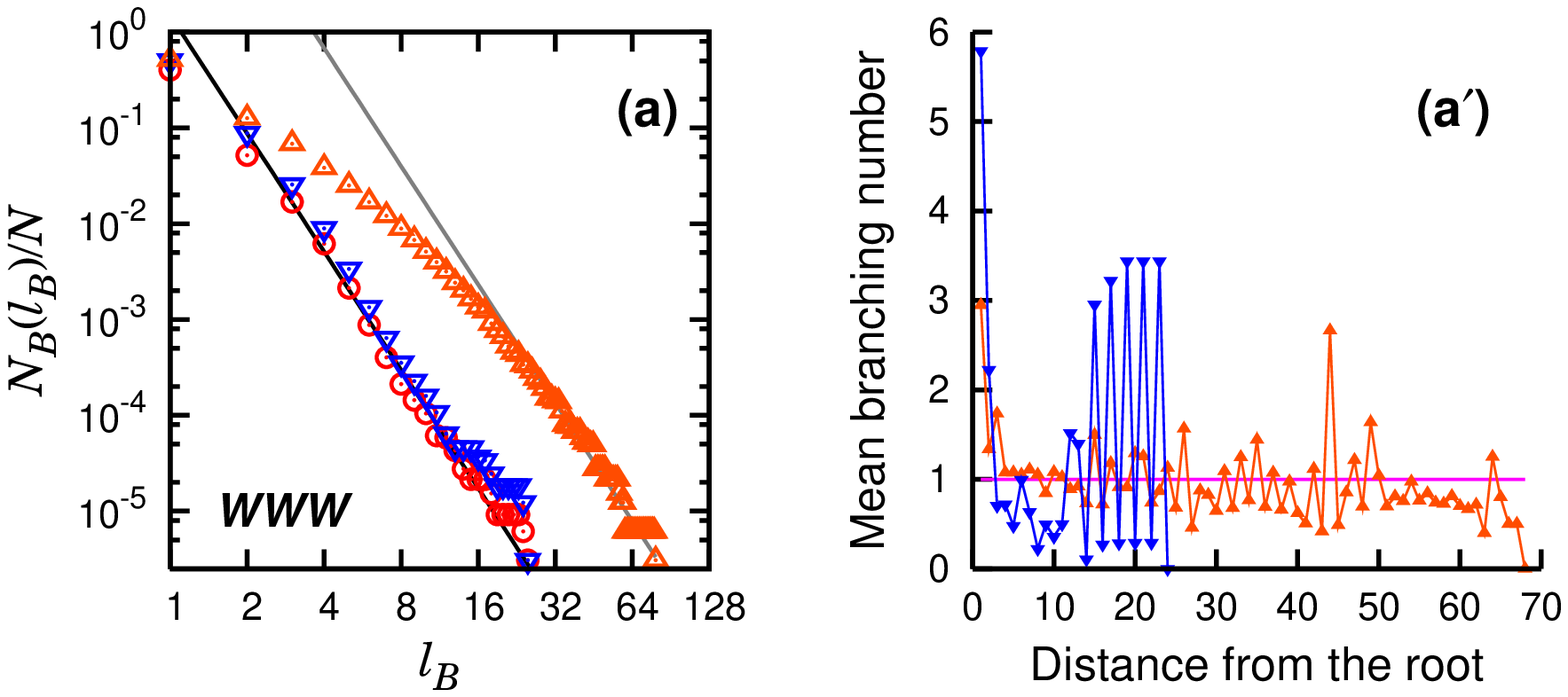}}
\centerline{\epsfxsize=9.5cm \epsfbox{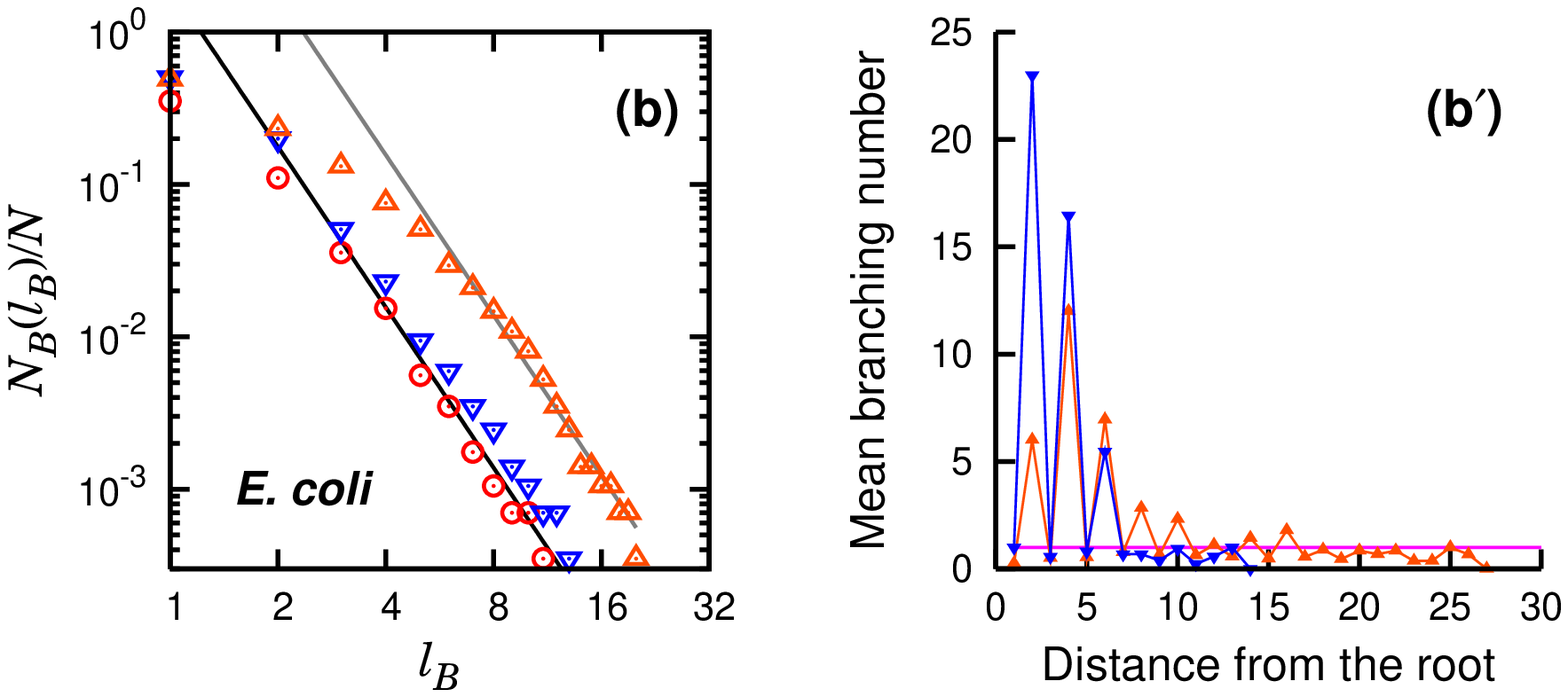}}
\centerline{\epsfxsize=9.5cm \epsfbox{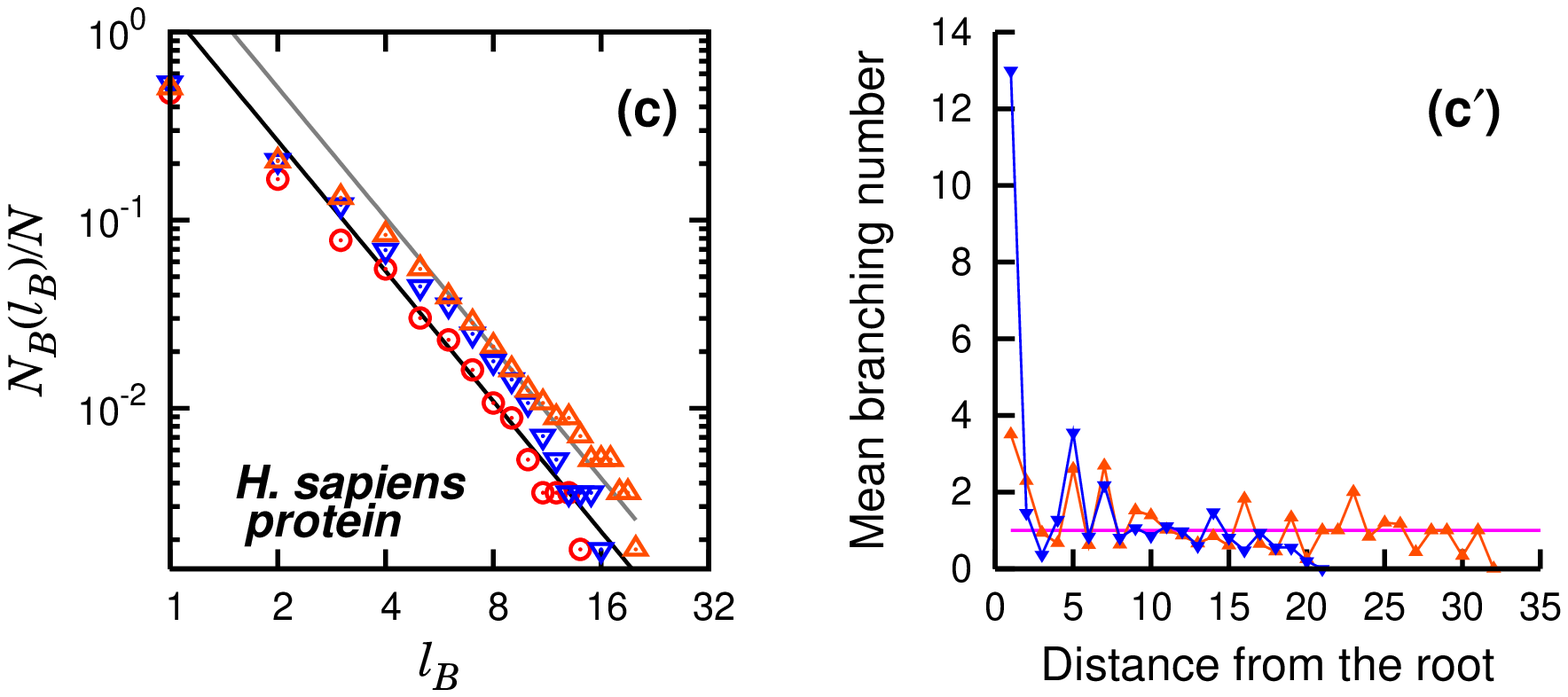}}
\centerline{\epsfxsize=9.5cm \epsfbox{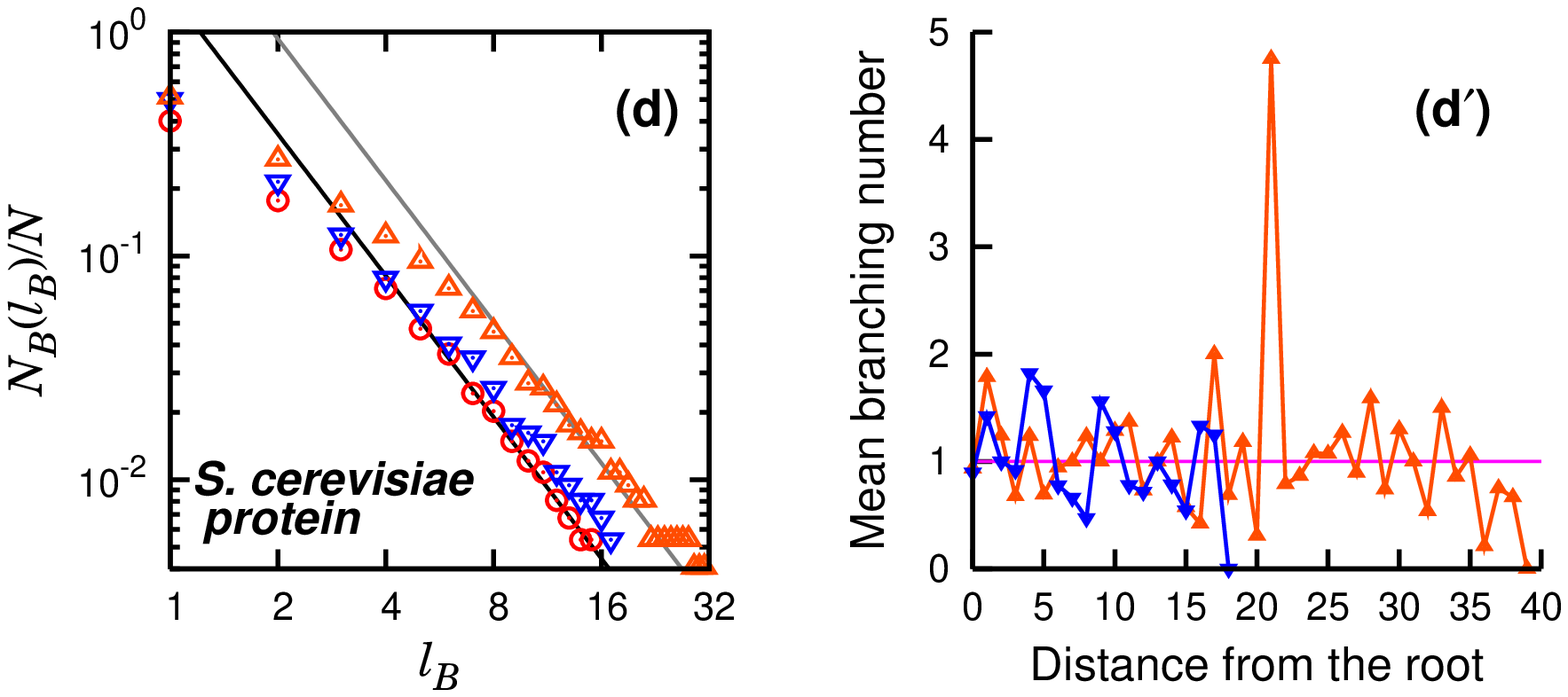}}
\centerline{\epsfxsize=9.5cm \epsfbox{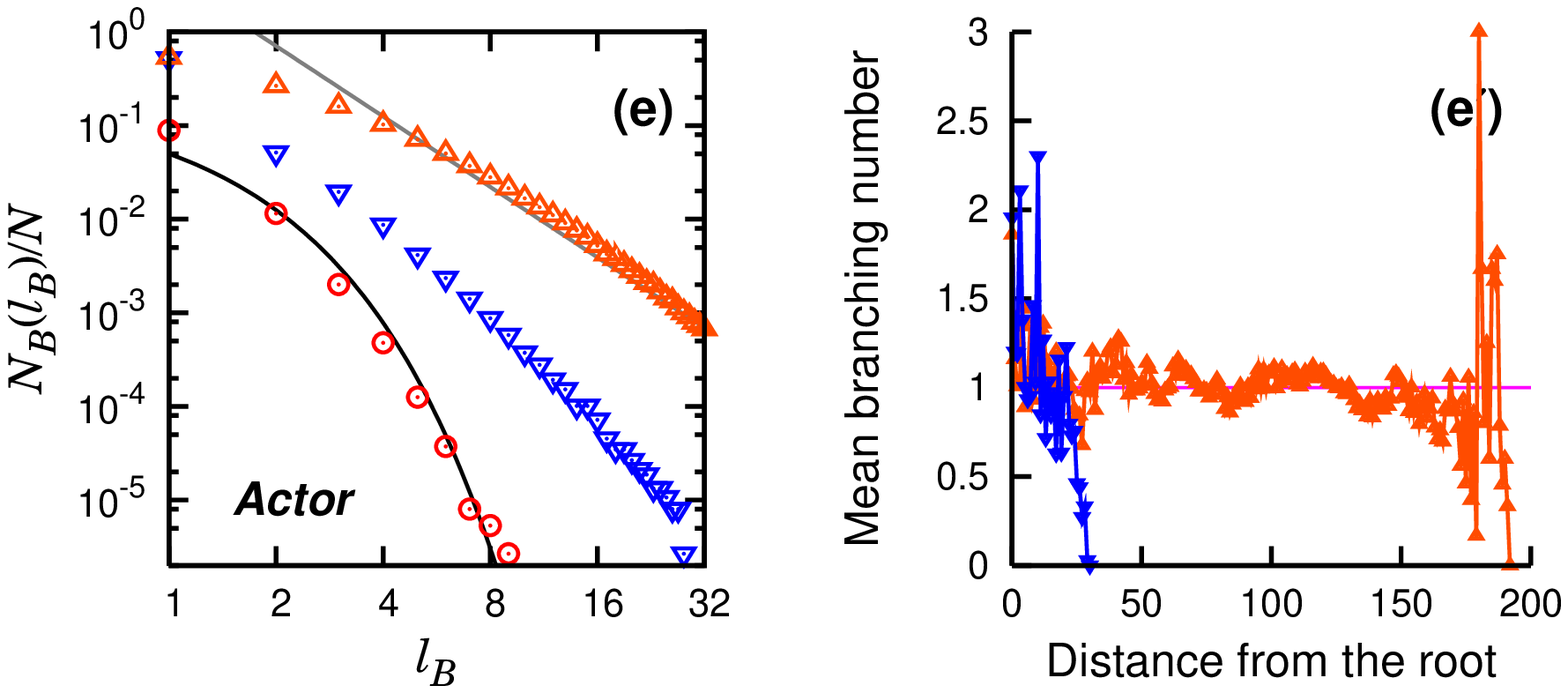}}
\end{minipage}\hfill
\begin{minipage}{.5\linewidth}
\centerline{\epsfxsize=9.5cm \epsfbox{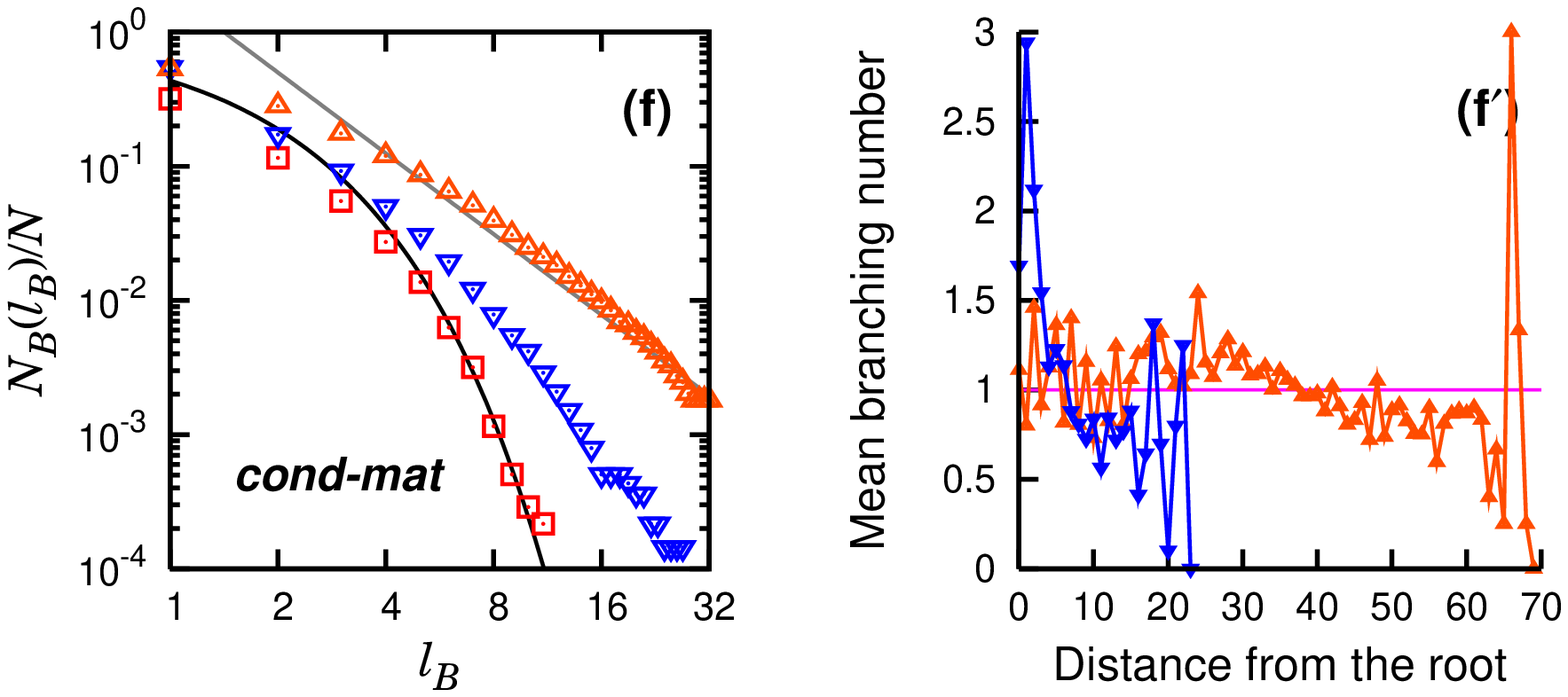}}
\centerline{\epsfxsize=9.5cm \epsfbox{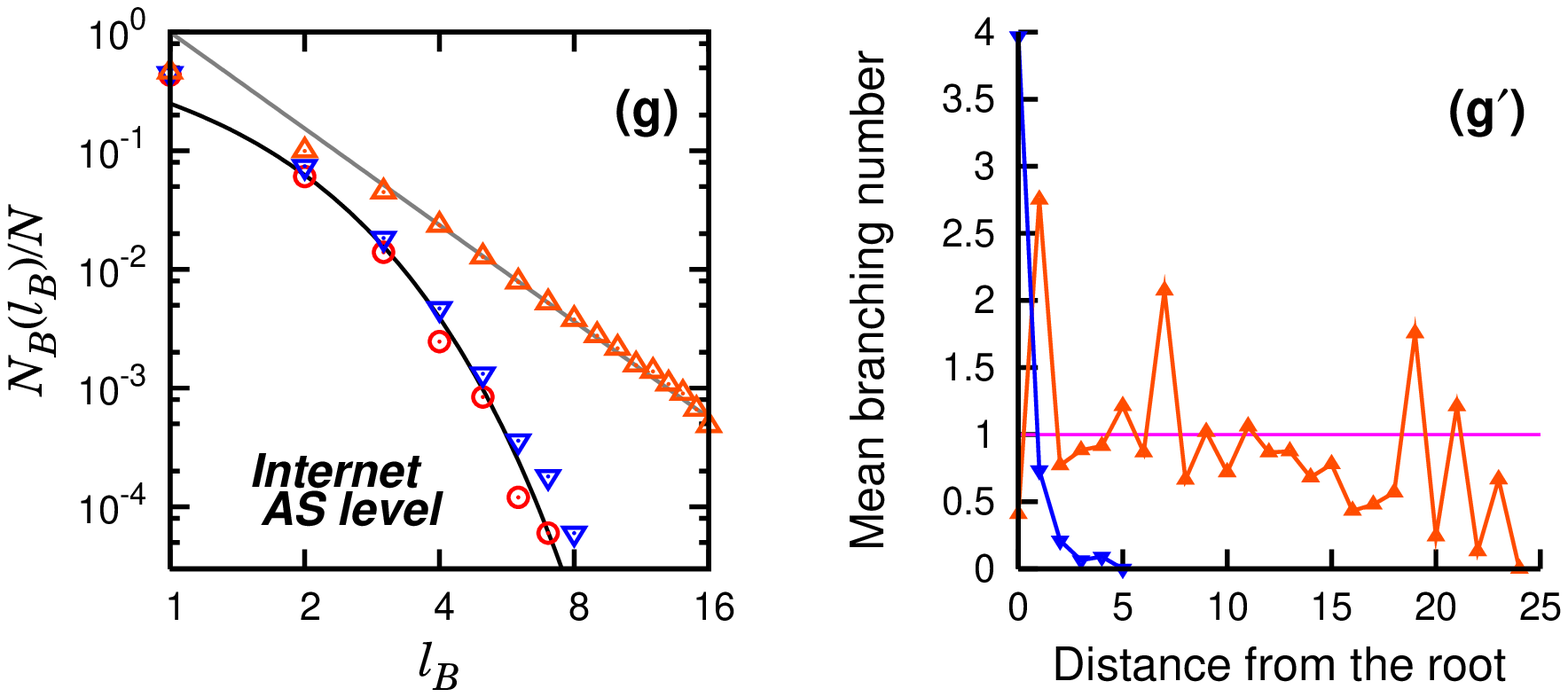}}
\centerline{\epsfxsize=9.5cm \epsfbox{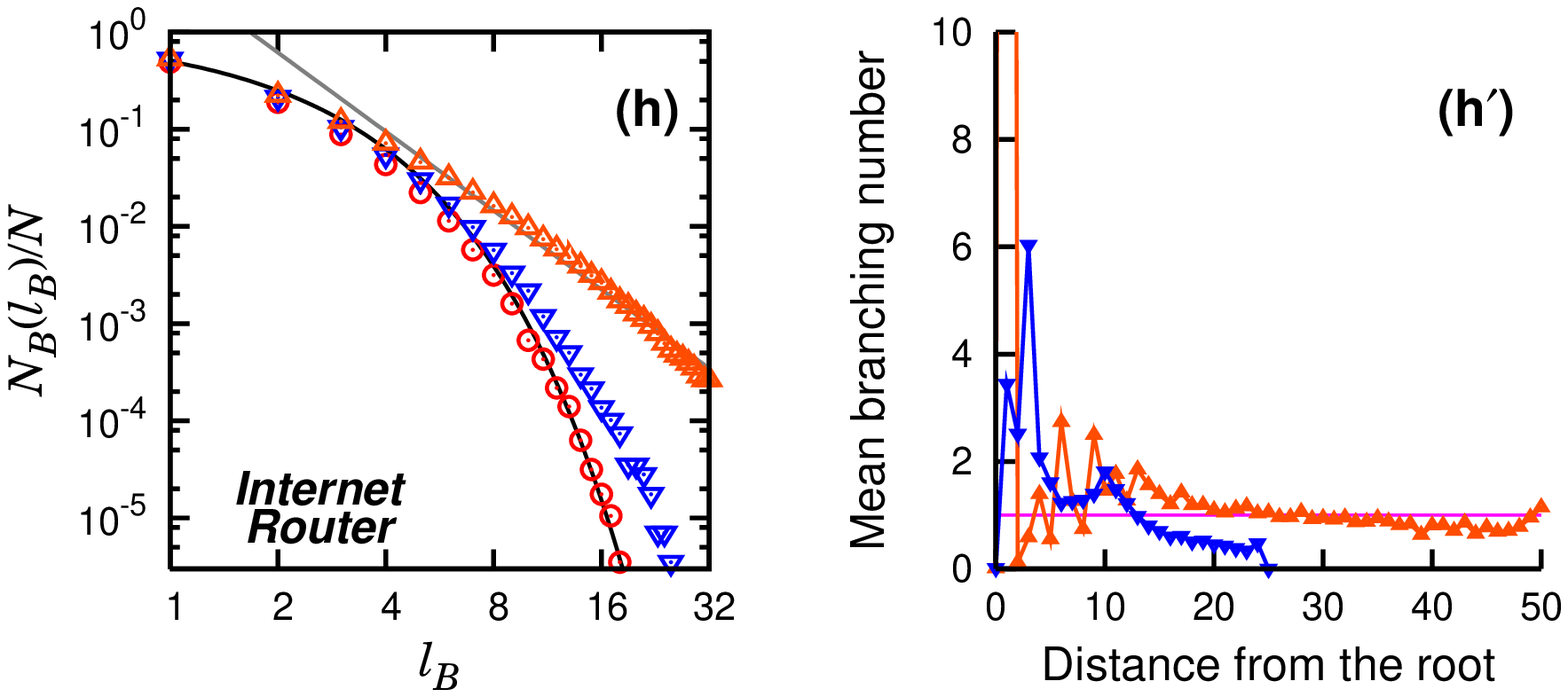}}
\centerline{\epsfxsize=9.5cm \epsfbox{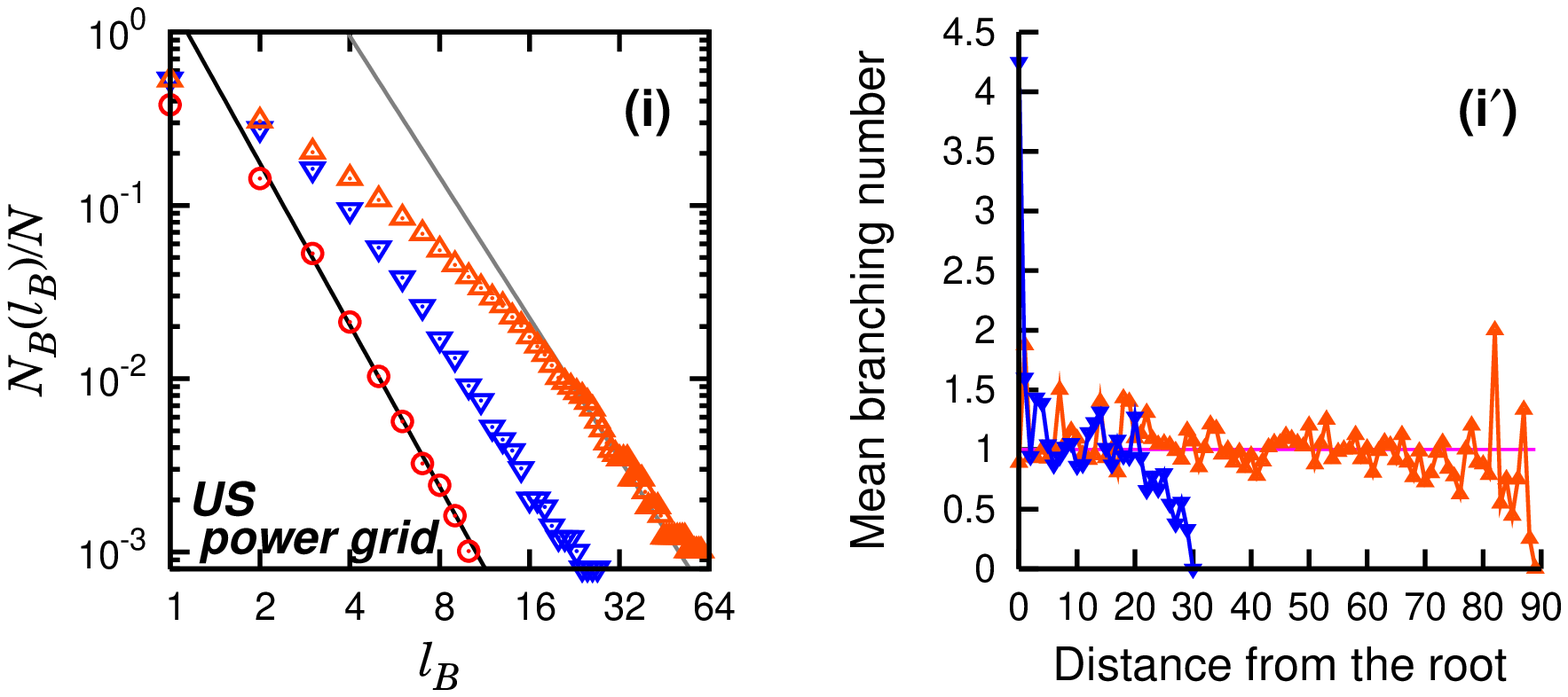}} \caption{Fractal
scaling analysis (left column) and mean branching number (right
column) of real-world networks including fractal (a)--(d) and
non-fractal (e)--(i) networks. For each network, the original
network (\textcolor{red}{$\circ$}), the skeleton
(\textcolor{blue}{$\triangledown$}), and a random spanning tree
(\textcolor{orange}{$\triangle$}) are studied. In (a)--(d), the
straight lines, drawn for guidance, have slopes of $-4.1$, $-3.5$,
$-2.3$, and $-2.1$, respectively. In (e)--(h), the fit for the
exponential function in the case of the original network and a
power-law fit for the random spanning tree are shown. In the right
panels (a$^{\prime}$)--(i$^{\prime}$), the horizontal line at 1 is
drawn for reference. }
\end{minipage}
\label{fig3}
\end{figure*}

We present fractal scaling analysis for real-world networks, which
are listed in the left column of Fig.~ref{fig3}. We first examine
fractal networks such as (a) the world-wide web (WWW), (b) the
metabolic network of {\it E. coli}, (c) protein interaction
network (PIN) of {\it H. sapiens}, and (d) of {\it S.
cerevisiae}~\cite{yeast}. Next, non-fractal networks such as (e)
the actor network~\cite{actor}, (f) coauthorship network
\cite{arxiv}, (g) Internet at the autonomous system (AS) level
\cite{routeviews}, (h) Internet at the router level \cite{scan},
and (i) power-grid of the USA~\cite{ws} are studied. The
characteristics of these networks are listed in Table I. Note that
in the previous study by Song {\it et al.}, the protein
interaction network of {\em S.~cerevisiae} was classified as a
non-fractal network. In this work, we use a different
dataset~\cite{yeast} of high-confidence protein interactions, for
which the PIN is a fractal network.

For the fractal networks (a)--(d), the original network and its
skeleton exhibit the same fractal scaling behavior, and the
respective statistics of the numbers of boxes needed to cover them
are almost identical as shown in the left column of
Figs.~3(a)--(d). The fractal dimensions for these networks are
measured to be $\approx 4.1$, 3.5, 2.3, and 2.1 for (a) the WWW,
(b) the metabolic network, PIN of (c) human, and that of (d)
yeast, respectively. A power-law behavior is not observed for
non-fractal networks, in which $N_B(\ell_B)$ decays faster than
any power law. We also study the fractal scaling for a randomly
spanning tree of each network, which is constructed from edges
that are randomly selected from the original network to form a
tree. Since edges are selected randomly, the degree distribution
of the original network is conserved in the random spanning tree
\cite{kertesz}. The random spanning tree is fractal irrespective
of the fractality of the original networks; this follows from the
percolation theory~\cite{optimal}. Thus, the random spanning tree
follows a power law in the fractal scaling.

For non-fractal networks, the box number $N_B$ does not follow a
power law with respect to the box lateral size $\ell_B$. The
statistics of box number of skeletons differ significantly from
those of their original network; however, there are some
exceptions in our examination. For the Internet at the AS level
[Fig.~\ref{fig3}(g)], $N_B$ for the original network and skeleton
exhibits similar behavior; however, they do not follow a power
law. On the other hand, in the power grid [Fig.~\ref{fig3}(i)],
the fractal scalings of the original network and the skeleton
exhibit the same power-law exponent; however, the box numbers for
the original network and the skeleton differ significantly.
Although the fractal scaling exhibits a power-law behavior, this
fractality is not obvious, because the network size is too small
for checking if a plateau is intrinsically formed in $\bar{n}(d)$.
Thus, the fractality cannot be classified.

\begin{figure*}
\begin{minipage}{.5\linewidth}
\centerline{\epsfxsize=9.5cm \epsfbox{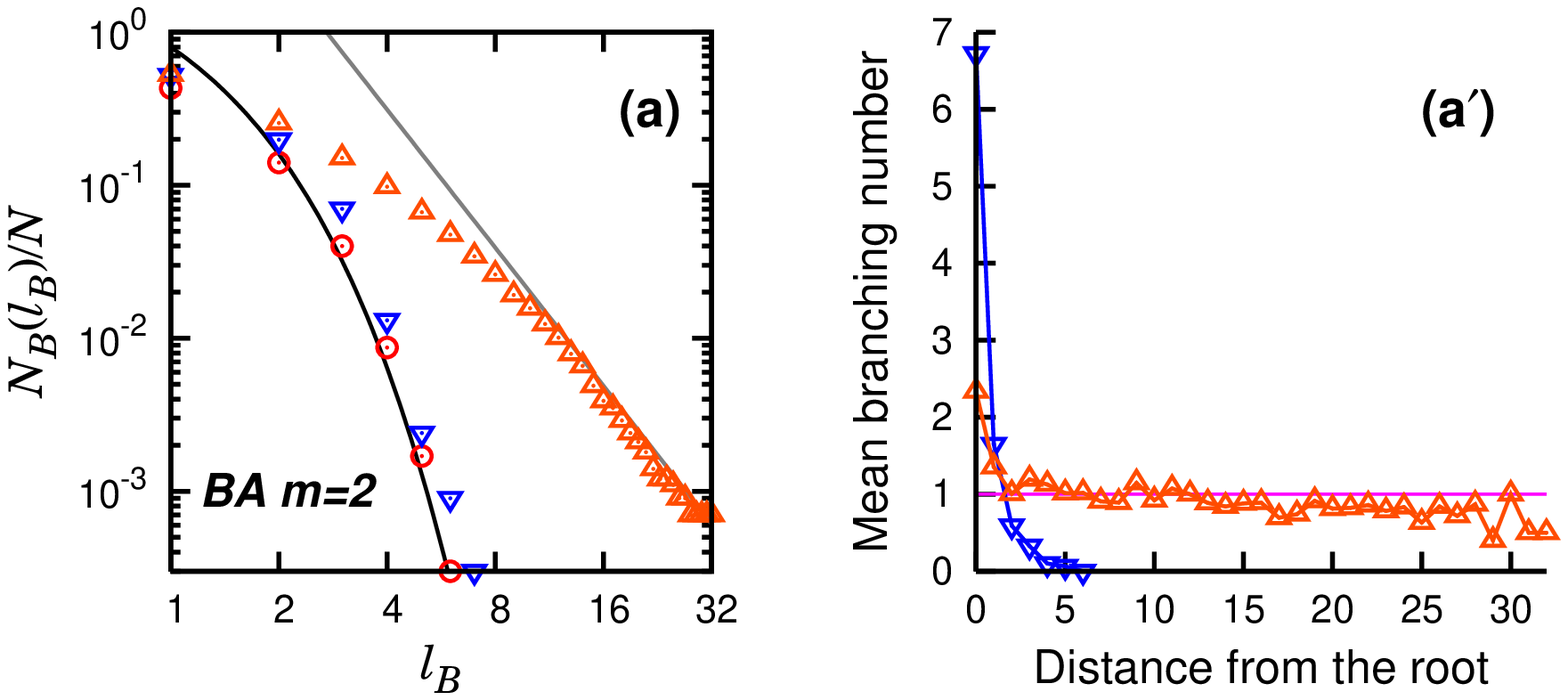}}
\centerline{\epsfxsize=9.5cm \epsfbox{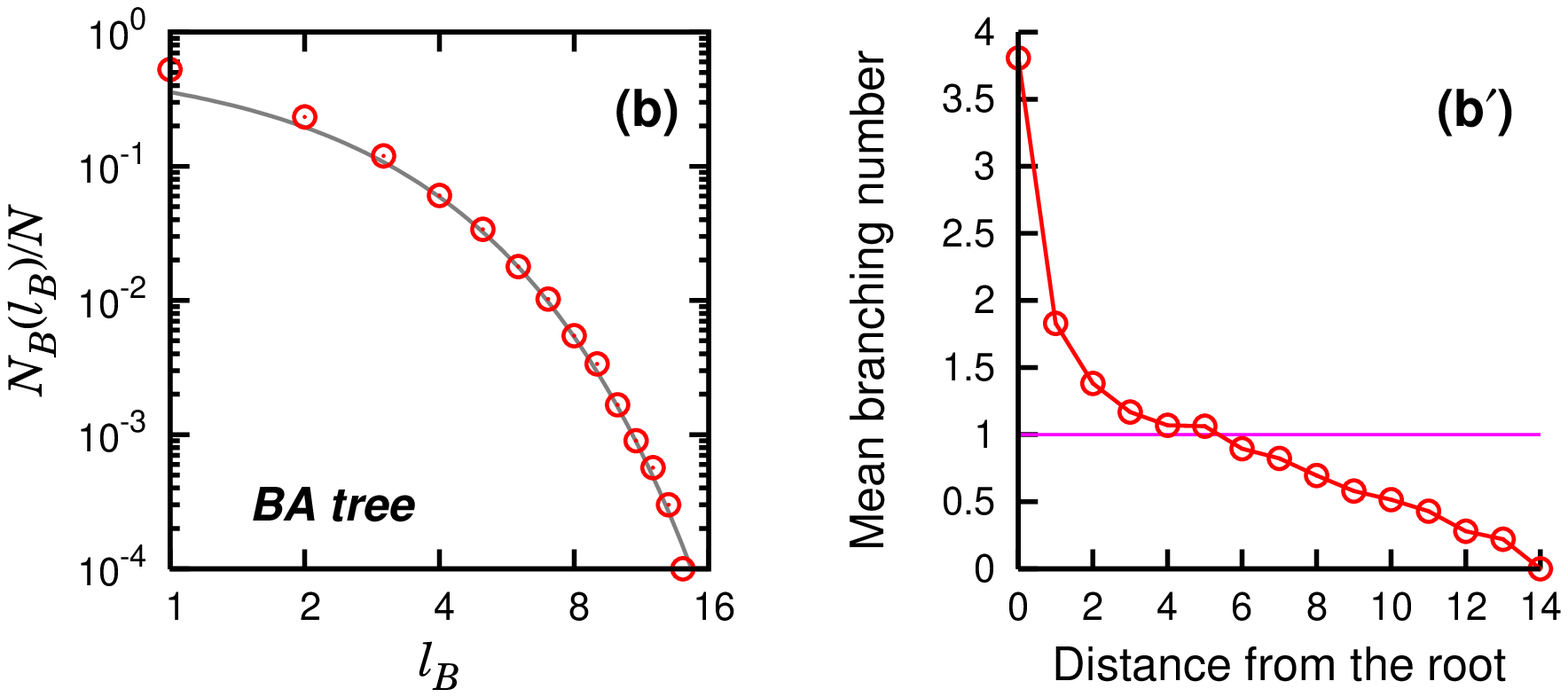}}
\centerline{\epsfxsize=9.5cm \epsfbox{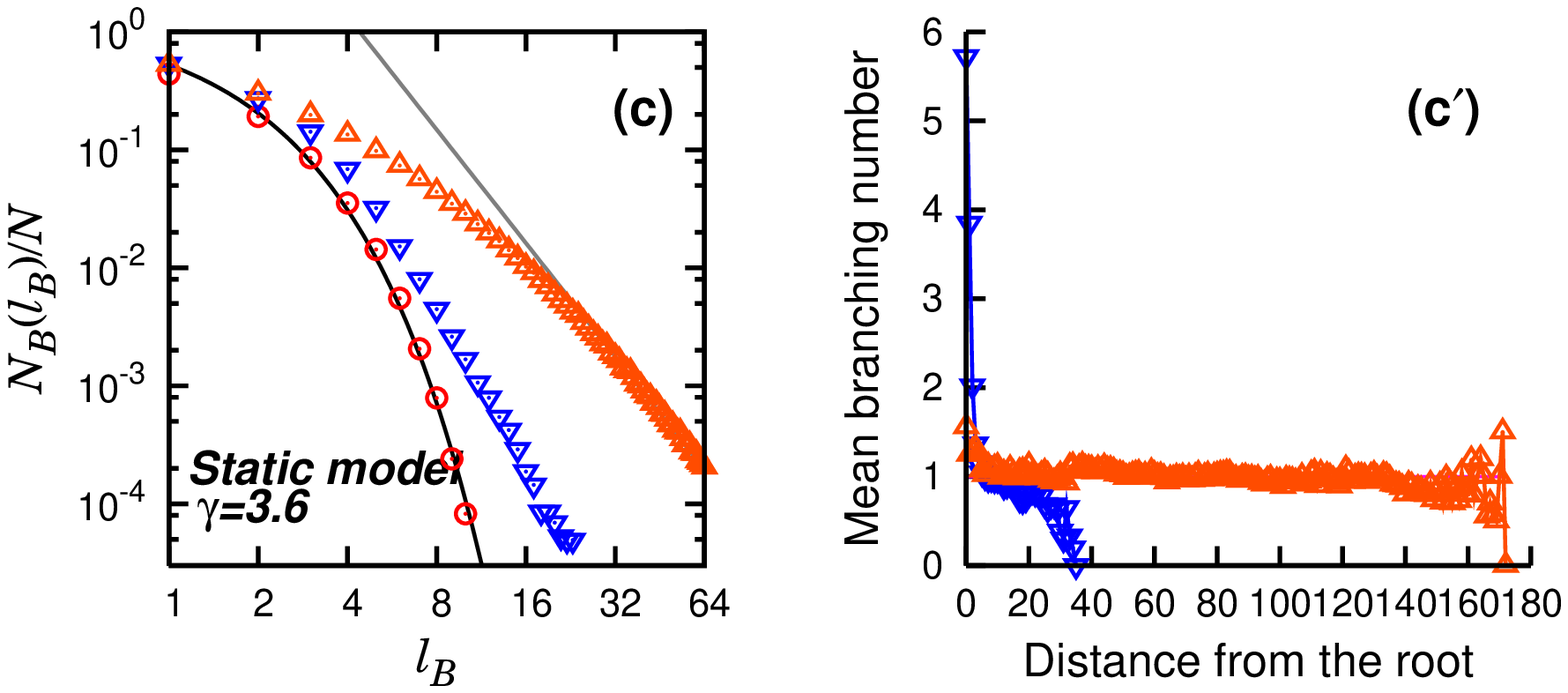}}
\end{minipage}\hfill
\begin{minipage}{.5\linewidth}
\centerline{\epsfxsize=9.5cm \epsfbox{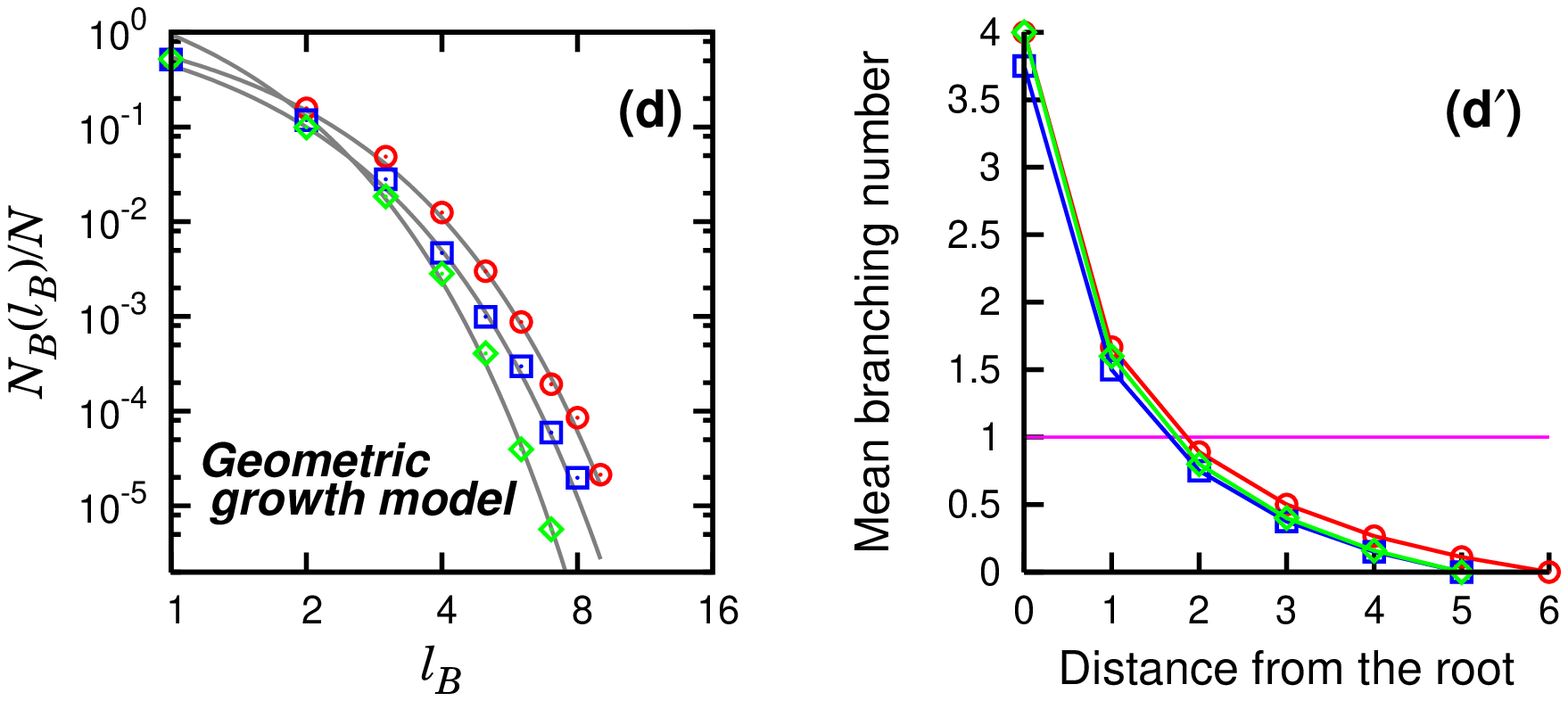}}
\centerline{\epsfxsize=9.5cm \epsfbox{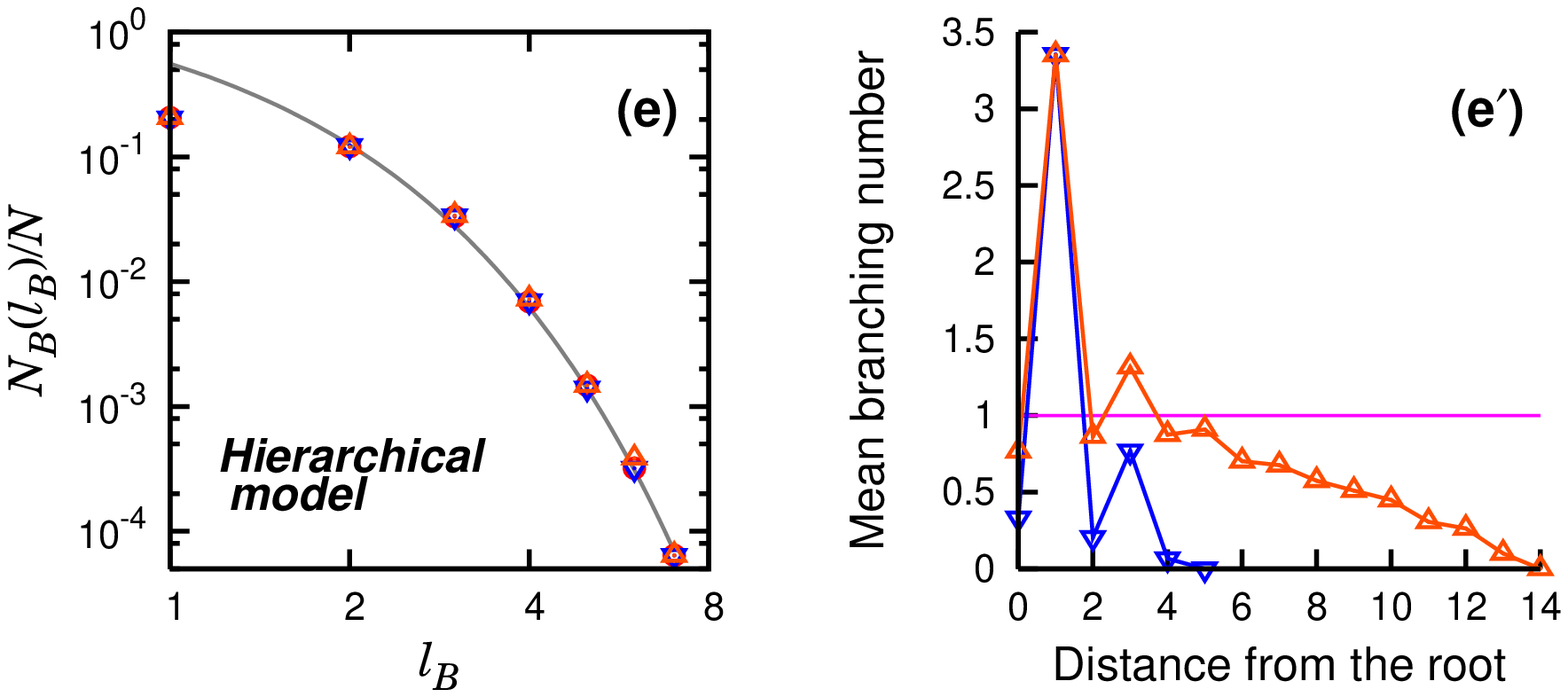}}
\caption{Fractal scaling analysis (a)--(e) and mean branching
number (a$^{\prime}$)--(e$^{\prime}$) of the SF network models.
The symbols used are similar to those in Fig.~\ref{fig3} except in
(d). The different symbols in (d) represents different parameters
in the model. In (a) and (c), the exponential fit for the original
network and a power-law fit for the random spanning tree are
shown. For (b) and (d), which are tree networks, the exponential
fit is shown. In (e), the exponential fit to the original network
is shown. Note that in (e), the box numbers for the original
network, the skeleton, and the random spanning tree all overlap.
In panels (a$^{\prime}$)--(e$^{\prime}$), a horizontal line at 1
is drawn for reference.} \end{minipage} \label{fig4}
\end{figure*}

Next, we examine fractal scaling in the following network models
(a) the Barab\'asi-Albert model with the degree of incident vertex
$m=2$ \cite{ba}, (b) Barab\'asi-Albert model tree with $m=1$, (c)
static model~\cite{static}, (d) geometric growth
model~\cite{jung}, and (e) deterministic hierarchical
model~\cite{hmodel} in Fig.~\ref{fig4}. The network models
considered do not obey the power-law fractal scaling; therefore,
they are not fractals.

\section{Mean branching number analysis}

We present the mean branching number (MBN) analysis for the
skeleton and random spanning trees of each network considered. We
define the MBN function $\bar{n}(d)$ as the mean number of
offsprings of each vertex at distance $d$ from the root in a
branching tree. For the fractal networks
[Figs.~3(a$^{\prime}$)--(d$^{\prime}$)], both the skeleton and the
random spanning tree exhibit a plateau in MBN, a signature of a
persistent branching structure. For random spanning trees, the
location of the plateau is distinctly obtained as
$\bar{n}\approx1$. With regard to the skeletons, while the
plateaus in MBN of the protein interaction networks
[Figs.~3(c$^{\prime}$)--(d$^{\prime}$)] appear to be located
around $\bar{n}\approx1$, they cannot be located clearly for the
WWW and metabolic network due to large fluctuations
[Figs.~3(a$^{\prime}$)--(b$^{\prime}$)]. Such fluctuations may
originate due to various factors such as the finite-size effect
and the artificial choice of the root of the branching tree. The
dynamic origin of the formation of real-world networks may well be
more complicated than the purely random branching dynamics: Thus,
nontrivial correlations may exist. Although the location of the
plateau in MBN cannot be clearly determined in some cases, its
presence is a distinct feature of the fractal networks and is
absent in non-fractal networks.

For non-fractal networks, the MBN of the skeleton decays to zero
without forming a plateau
[Figs.~3(e$^{\prime}$)--(h$^{\prime}$))]. This is because the
skeleton of each non-fractal SF network belongs to the class of
``causal'' trees \cite{causal}, where vertices closer to the root
are likely to have larger degrees. In such structures, MBN
decreases steadily with the distance from the root; therefore, a
plateau cannot be formed. This absence of a plateau in MBN is also
observed in the skeletons of the network models shown in
Fig.~\ref{fig4}. Note that even for non-fractal networks, the
random spanning trees exhibit plateaus at $\bar{n}\approx1$,
confirming their fractality independent of the underlying original
network structure.

\begin{figure}
\begin{minipage}{8.5cm}
\begin{minipage}{0.495\linewidth}
{\epsfxsize=\linewidth \epsfbox{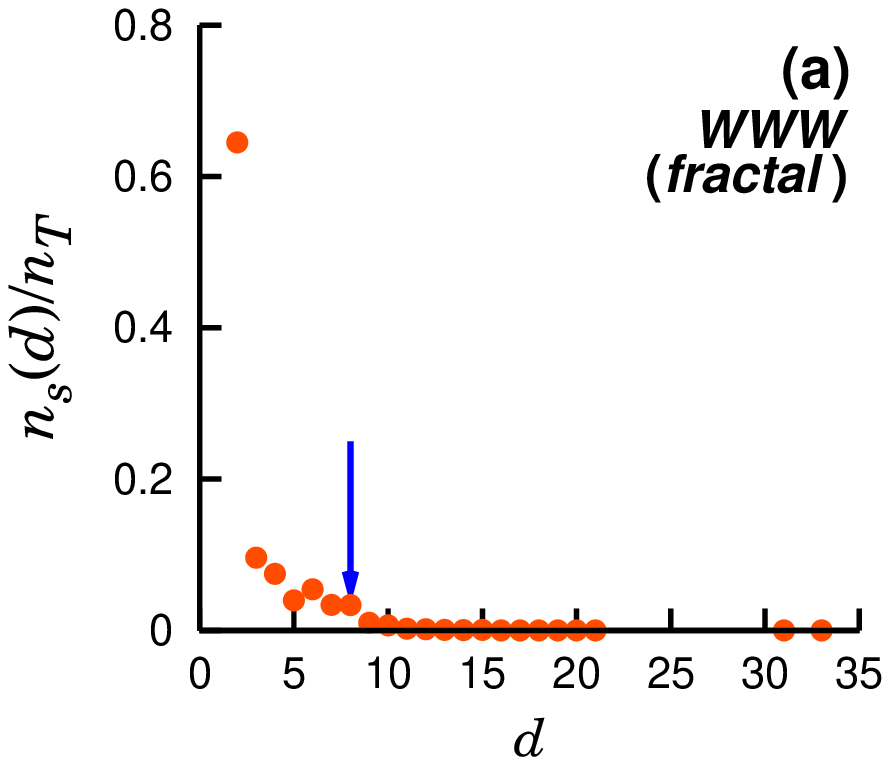}}
\end{minipage}\hfill
\begin{minipage}{0.495\linewidth}
{\epsfxsize=\linewidth \epsfbox{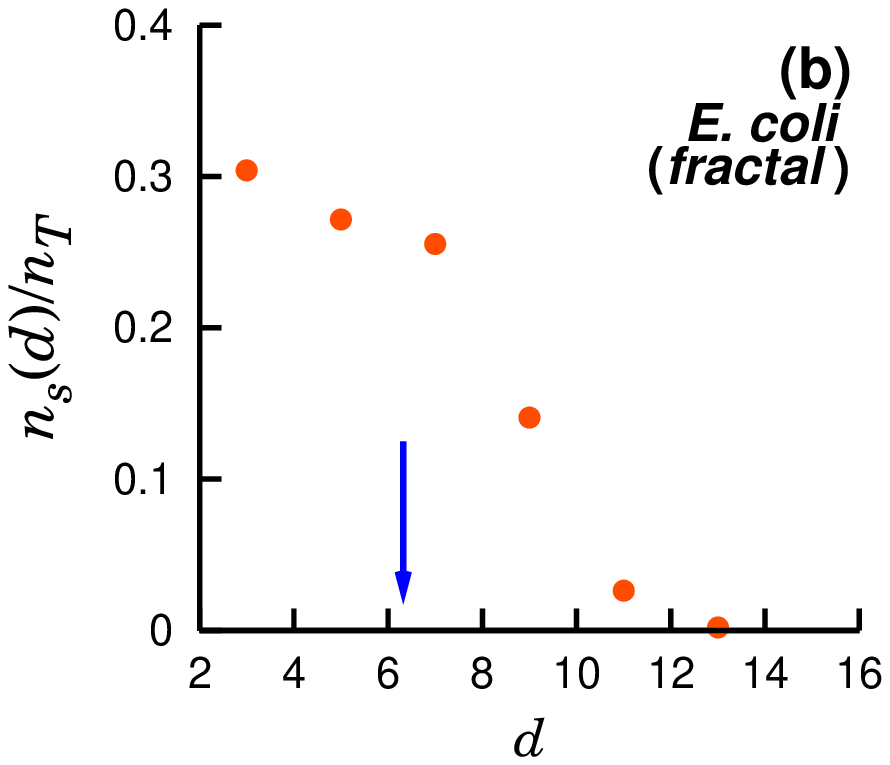}}
\end{minipage}
\begin{minipage}{0.495\linewidth}
{\epsfxsize=\linewidth \epsfbox{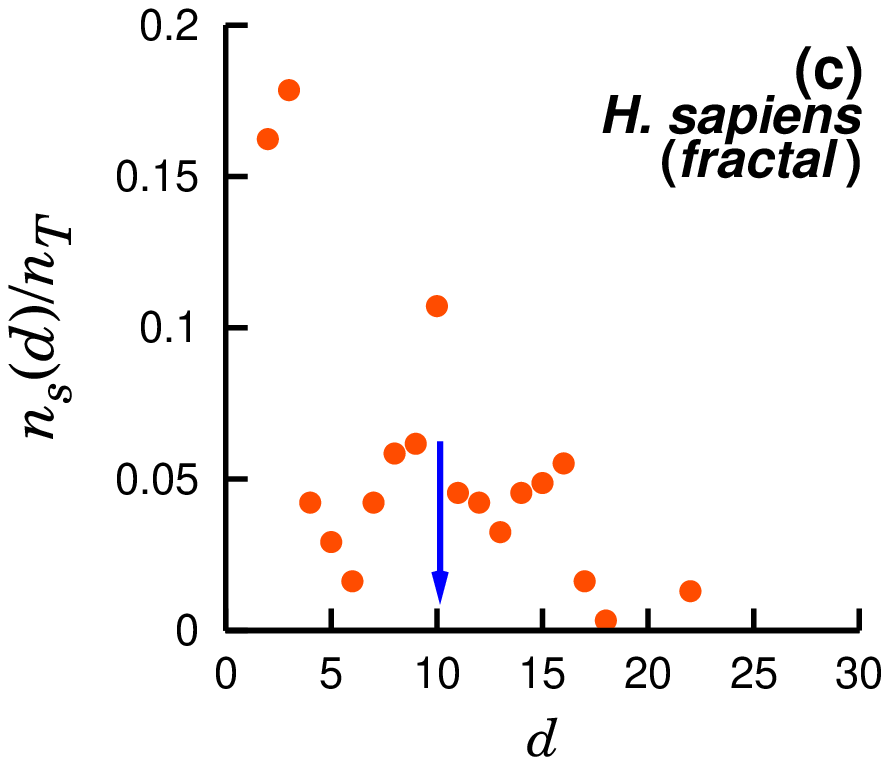}}
\end{minipage}\hfill
\begin{minipage}{0.495\linewidth}
{\epsfxsize=\linewidth \epsfbox{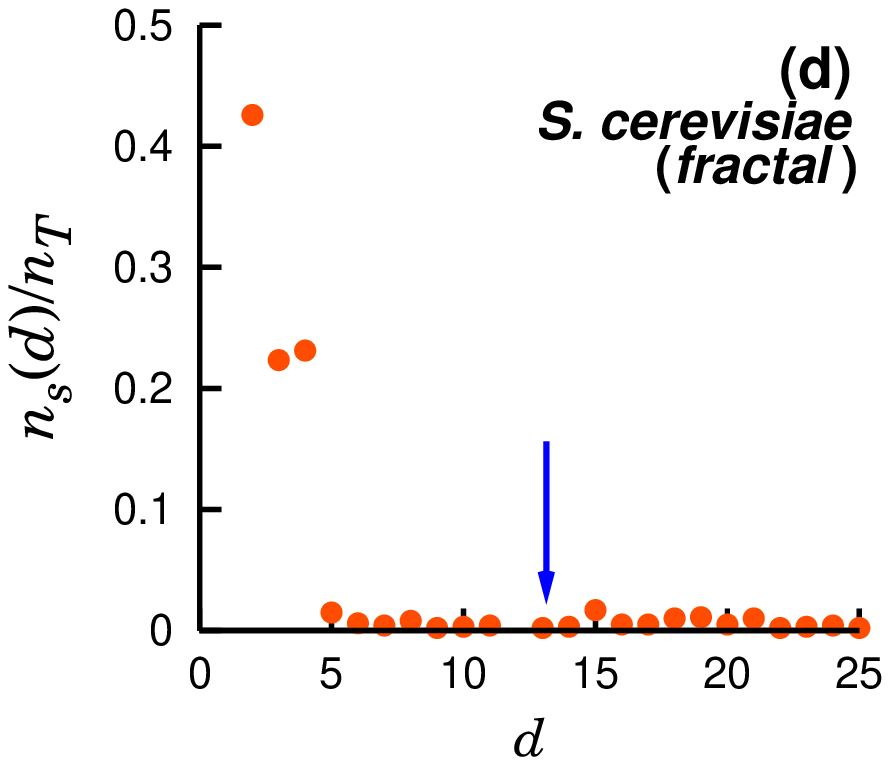}}
\end{minipage}
\begin{minipage}{0.495\linewidth}
{\epsfxsize=\linewidth \epsfbox{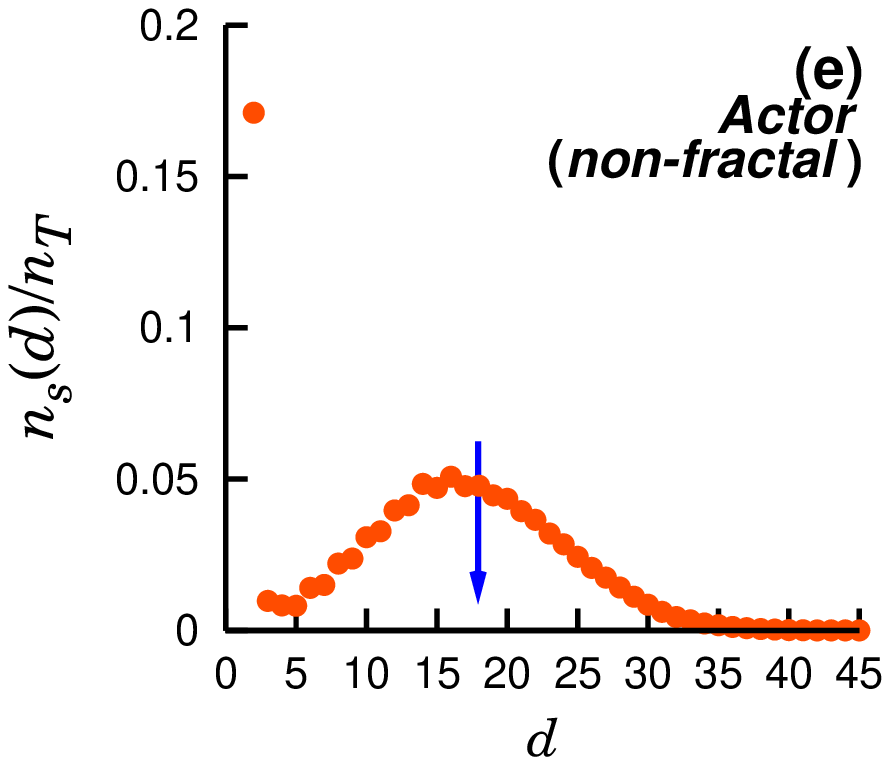}}
\end{minipage}\hfill
\begin{minipage}{0.495\linewidth}
{\epsfxsize=\linewidth \epsfbox{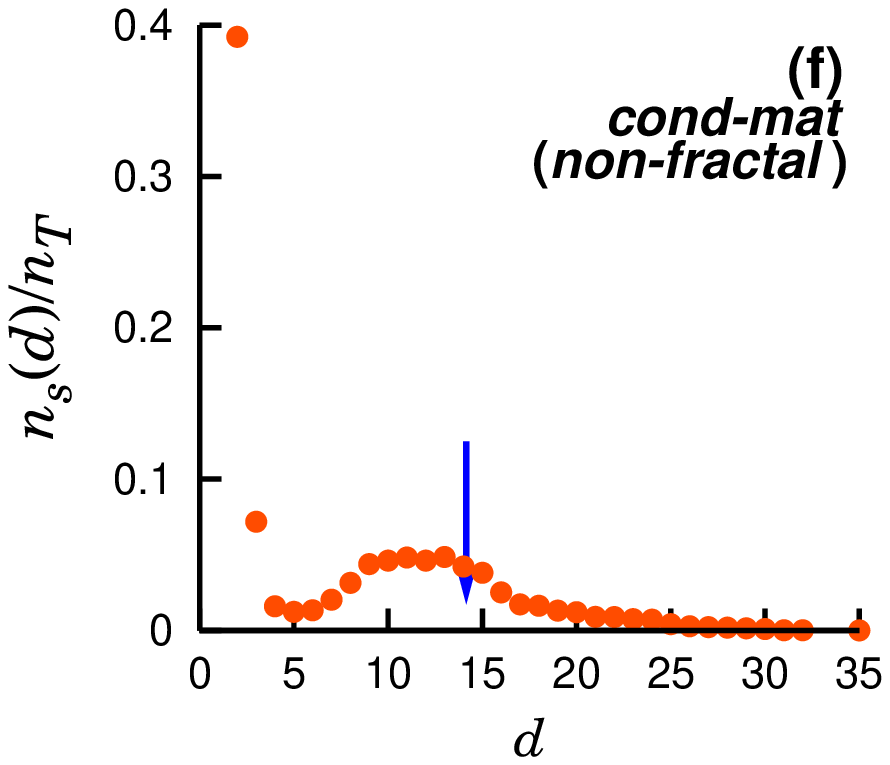}}
\end{minipage}
\begin{minipage}{0.495\linewidth}
{\epsfxsize=\linewidth \epsfbox{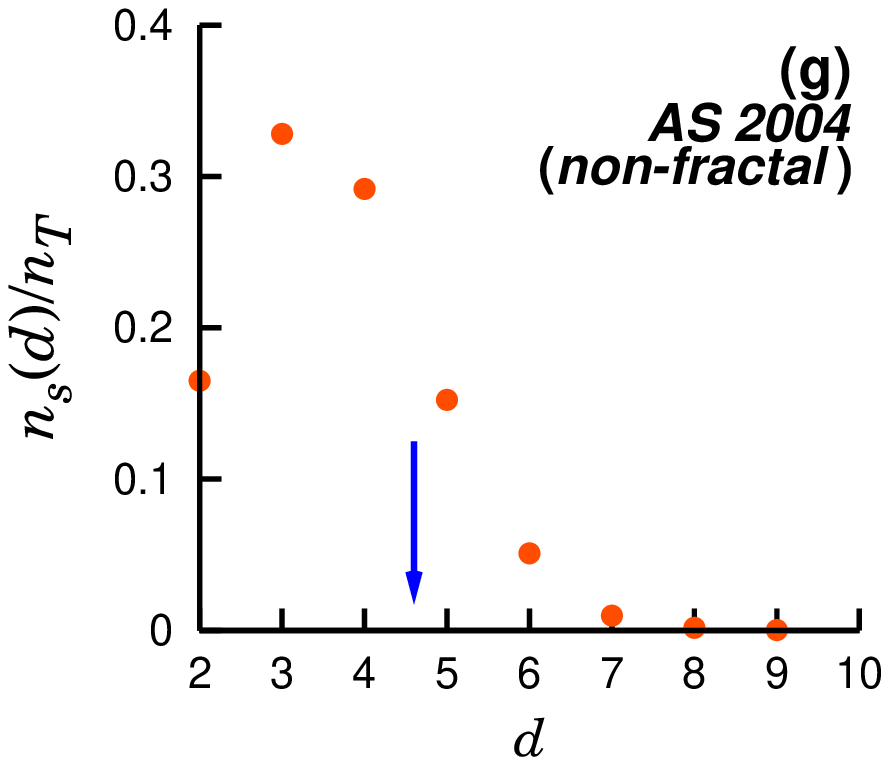}}
\end{minipage}\hfill
\begin{minipage}{0.495\linewidth}
{\epsfxsize=\linewidth \epsfbox{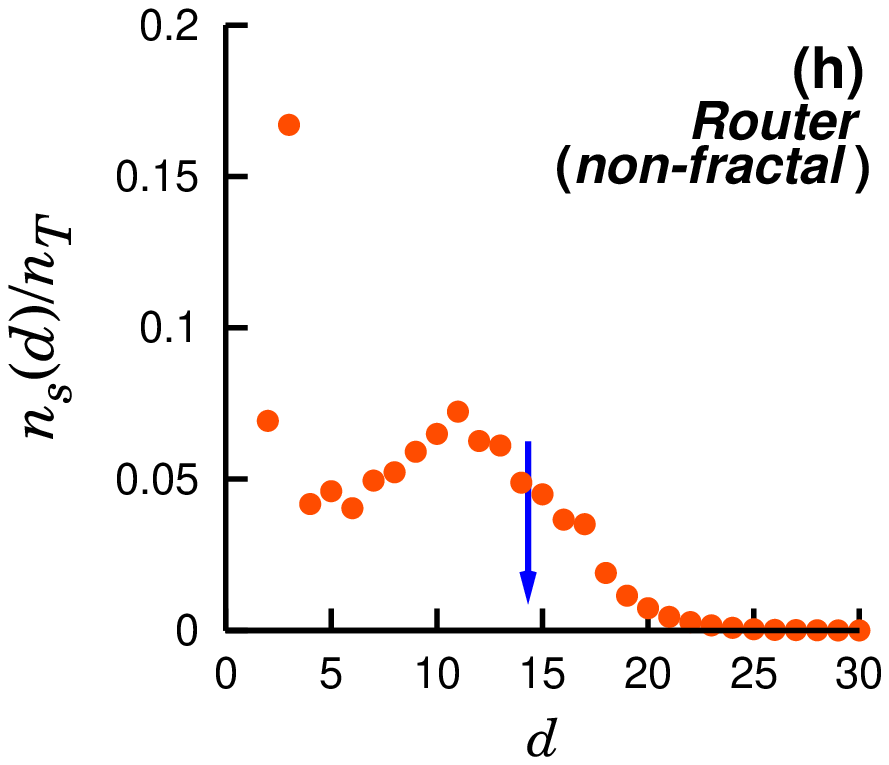}}
\end{minipage}
\begin{minipage}{0.495\linewidth}
{\epsfxsize=\linewidth \epsfbox{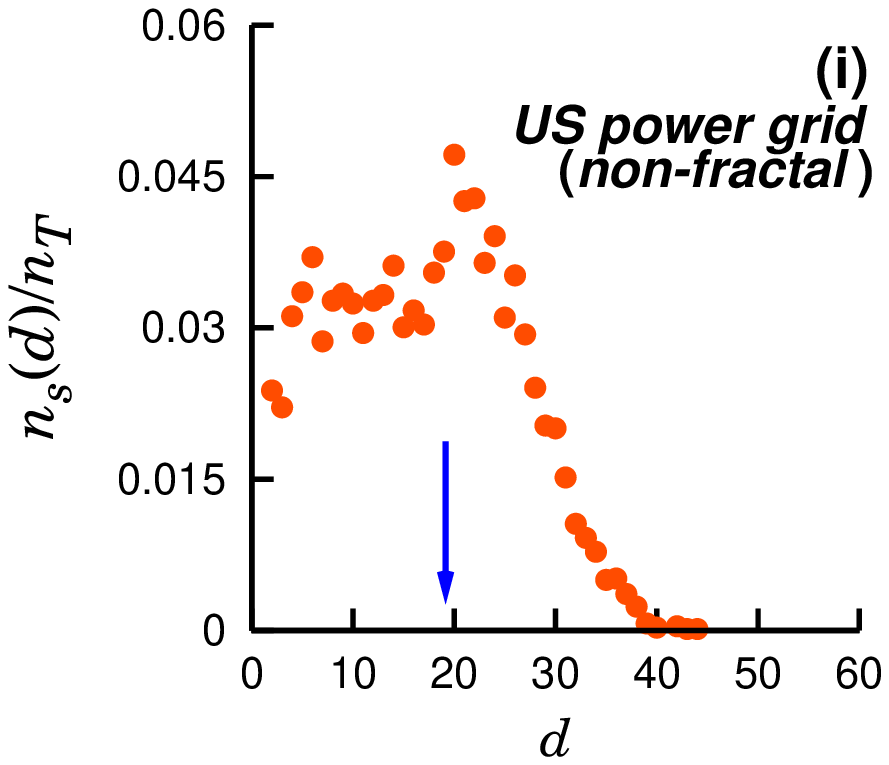}}
\end{minipage}\hfill
\begin{minipage}{0.495\linewidth}
{\epsfxsize=\linewidth \epsfbox{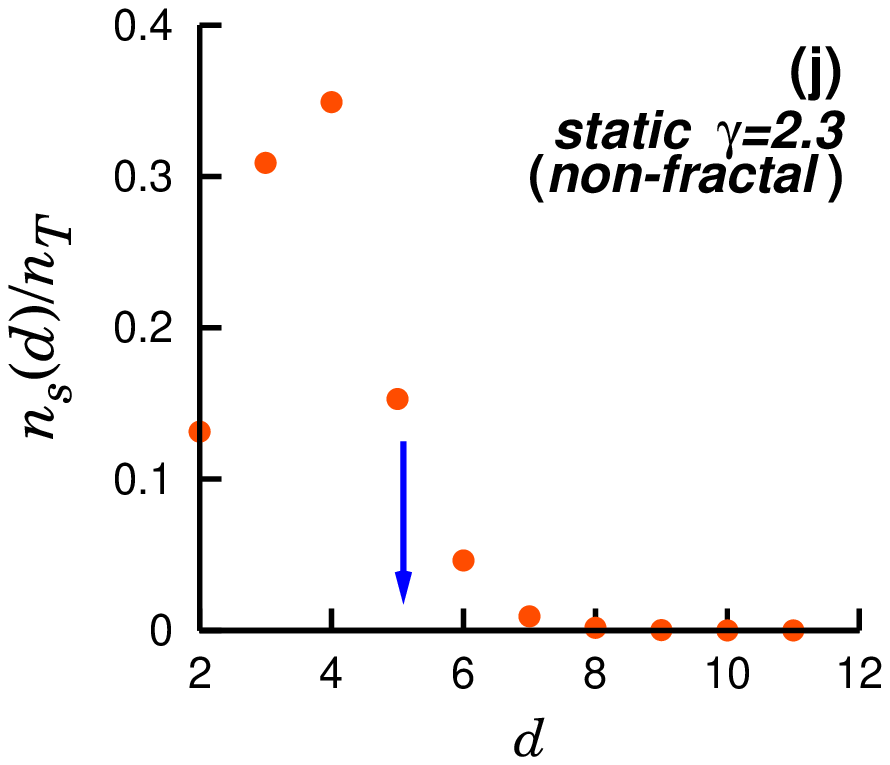}}
\end{minipage}
\end{minipage}
\caption{Length distribution of shortcuts. The length of a
shortcut is defined as the shortest distance along the skeleton
between the two vertices connected by the shortcut. The arrows
indicate the diameter of the skeleton of each network. Here,
$n_s(d)$ and $n_T$ are the number of shortcuts with length $d$ and
the total number of shortcuts, respectively.} \label{fig5}
\end{figure}

Although the fractal skeleton provides a scaffold for fractality
in fractal networks, the manner in which the shortcuts are placed
in the network is also important for preserving the fractality.
With regard to this, the previous result of the length
distribution of shortcuts~\cite{skeleton} is important. It is
known that two types of shortcut length distributions exist
\cite{skeleton}. In the first type, the shortcut length
distribution decays completely with respect to the shortcut
length. In the other type, the shortcut length distribution
exhibits a peak at a finite length, which is comparable to the
average separation of all pairs of vertices on each skeleton; this
is indicated by an arrow for each network in Fig.~\ref{fig5}.
Thus, in the former case, the shortcuts connecting different
branches of the skeleton are rare; however, their contribution is
considerable in the latter case. In the latter case, the network
will be globally interwoven and loses the fractality as in the
case of random SF networks. Indeed, we have found that the fractal
networks exhibit the former behavior [Figs.~\ref{fig5}(a)--(d)],
while the non-fractal networks exhibit the latter
[Figs.~\ref{fig5}(e)--(j)]. This indicates that the shortcuts in
the fractal networks are mainly local.

\section{Fractal network model}

The observation of a plateau in the MBN for the skeleton of the
fractal networks prompted the construction of a fractal network
model based on a random branching tree. We construct the model by
reversing the steps, followed thus far to reveal the fractality.
We first construct a branching tree in which the branching
proceeds stochastically with a prescribed branching probability
$b_n$. We choose $b_n$ to follow a power law with respect to $n$
in order to generate an SF network. Then, the branching tree is
dressed with local shortcuts as well as global ones. The global
connection is introduced to observe the crossover from fractal to
non-fractal behavior. The frequency of global shortcuts is an
important parameter of the model. More specifically, we consider
the branching probability $b_n$, i.e., the probability to generate
$n$ offsprings in each branching step, as \be b_n =
\frac{1}{Z}n^{-\gamma} \quad (\gamma>2) \ee for $n\ge 1$, and  \be
b_0=1-\sum_{n=1}^{\infty}b_n \ee for $n=0$. Then, the resulting
tree network is an SF network with the degree exponent $\gamma$.
In order to generate a {\em critical} branching tree, the
normalization constant $Z$ is set to be $Z=\zeta(\gamma-1)$.
Where, $\zeta(x)$ is the Riemann zeta function, which follows from
the criticality condition $\langle n\rangle=\sum_n nb_n=1$. We can
also generate a {\em supercritical} branching tree by setting
$Z=\zeta(\gamma-1)/\langle n \rangle$ with $\langle n\rangle>1$.

After we generate a branching tree, we dress it with shortcuts by
increasing the degree of each vertex by a factor $p$ and
establishing the available connections between vertices. This can
be achieved either in a local or global manner.  An additional
parameter $q$ is introduced to describe the frequency of global
shortcuts in the network. The creation rule of the fractal network
model is described as follows:
\begin{enumerate}
\item[{(i)}] We start with a seed vertex from which $n$
offsprings are stochastically generated with probability $b_n$
($n=0,...,N-1$). Each offspring then generates $n$ branches with
probability $b_n$. This process is repeated until we obtain a
network of desired size $N$. If the growth of the tree stops
before attaining size $N$, we restart the branching procedure.

\item[{(ii)}]
Degree $k_i$ of each vertex $i$ is increased by a factor $p$ such
that vertex $i$ obtains additional $p k_i$ stubs for forming
edges. From these stubs, $qpk_i$ stubs are assigned to global
shortcuts, while the remaining $(1-q)pk_i$ stubs are assigned to
local shortcuts. In order to establish local shortcuts, we search
vertices from the root. A vertex $i$ that has at least one stub
for local shortcuts is selected. Then, its connection partner is
scratched from the closest vertices from the vertex $i$ to a
vertex $j$, having available stubs for local shortcuts and not yet
connected to $i$, to form an edge between $i$ and $j$. This
process is repeated until all the stubs for local shortcuts are
linked.

\item[{(iii)}] Next, we choose two vertices
$i$ and $j$ randomly; each of which has at least one stub for
global shortcuts. We then connect them to form an edge if they are
not already connected. This process is repeated until all stubs
for global shortcuts are linked. This step is similar to the
process used in the configuration model~\cite{molloy}.
\end{enumerate}

\begin{figure*}
\begin{minipage}{18cm}
\begin{minipage}{0.280\linewidth}
{\epsfxsize=\linewidth \epsfbox{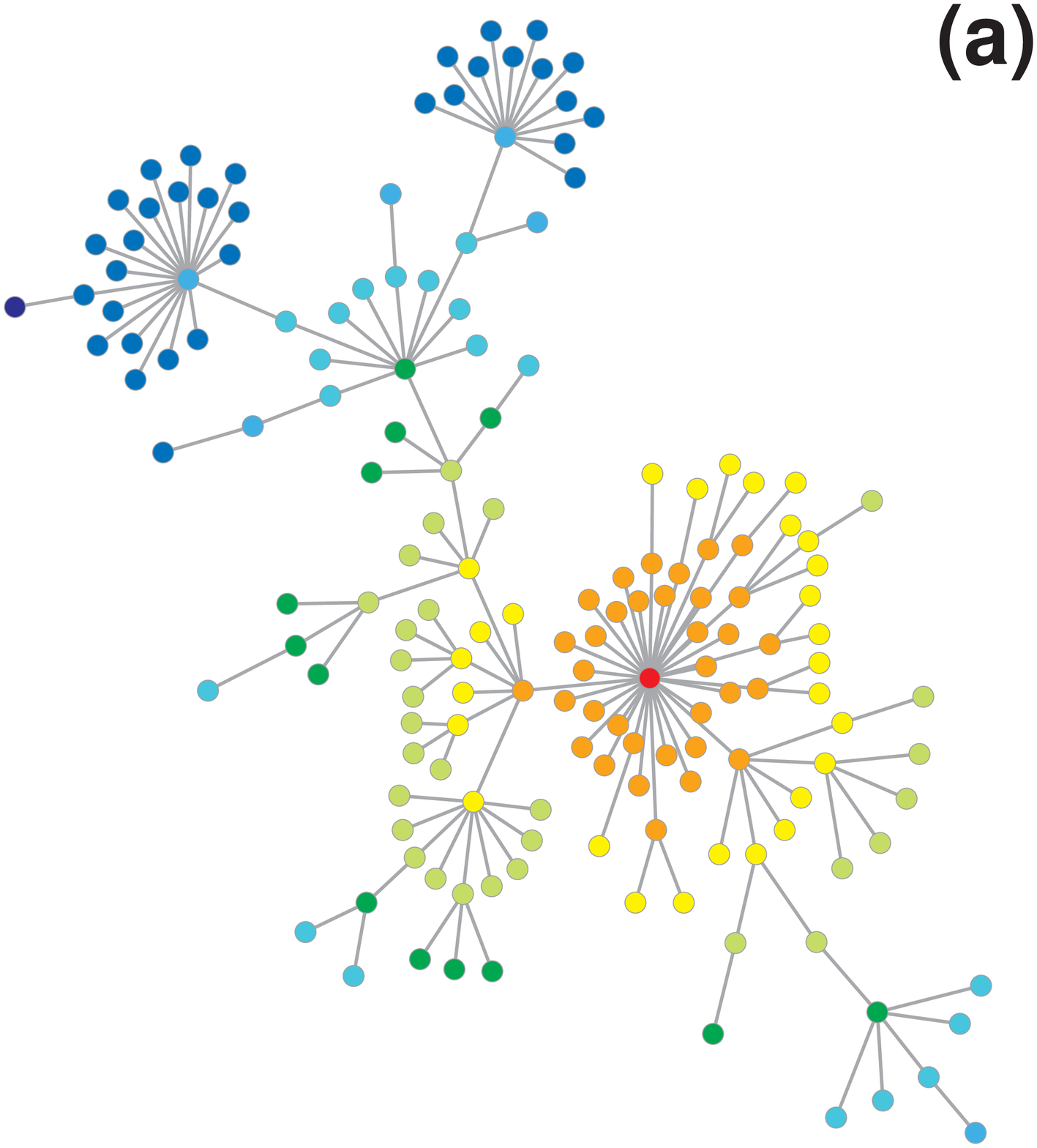}}
\end{minipage}
\begin{minipage}{0.280\linewidth}
{\epsfxsize=\linewidth \epsfbox{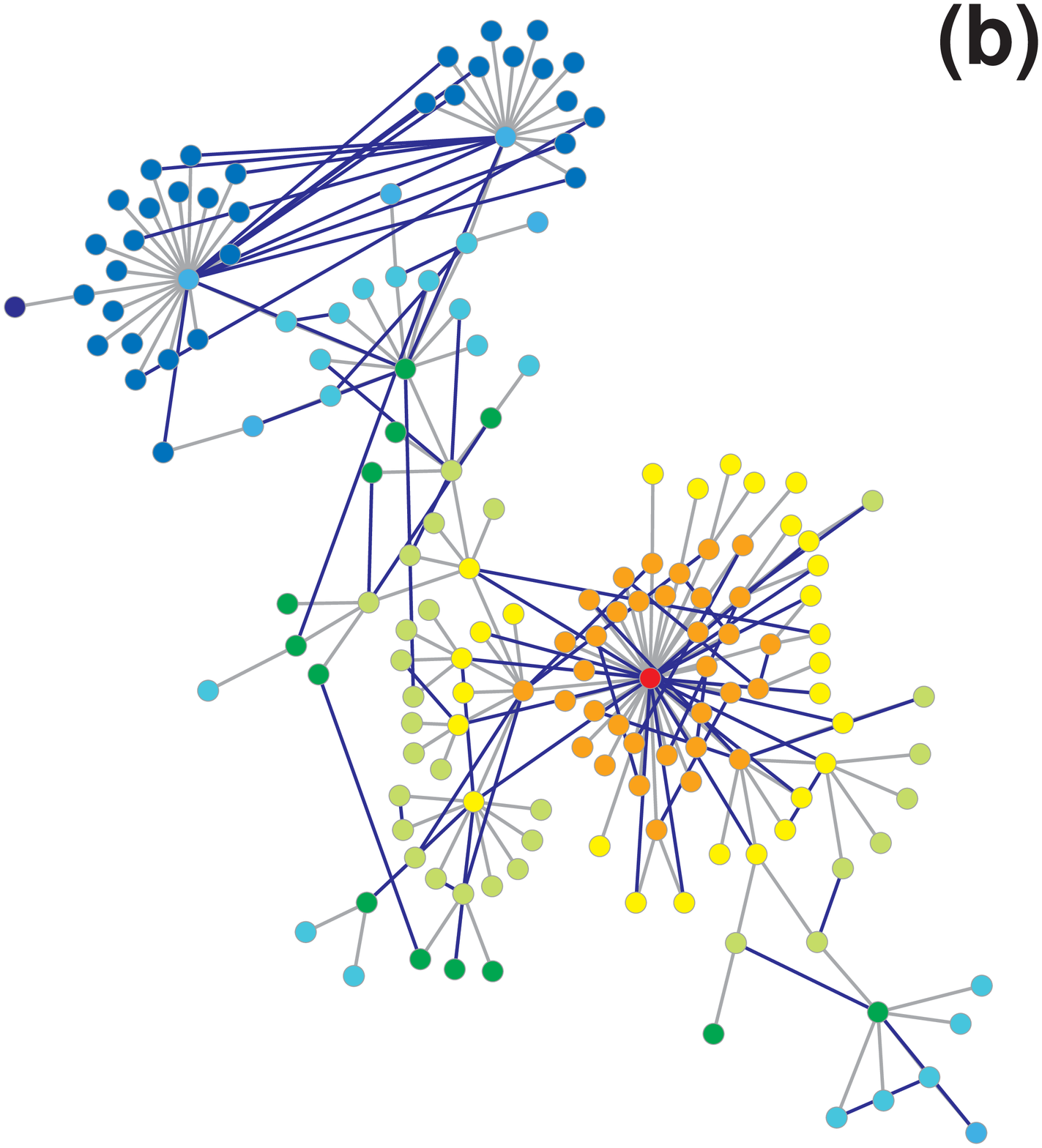}}
\end{minipage}
\begin{minipage}{0.280\linewidth}
{\epsfxsize=\linewidth \epsfbox{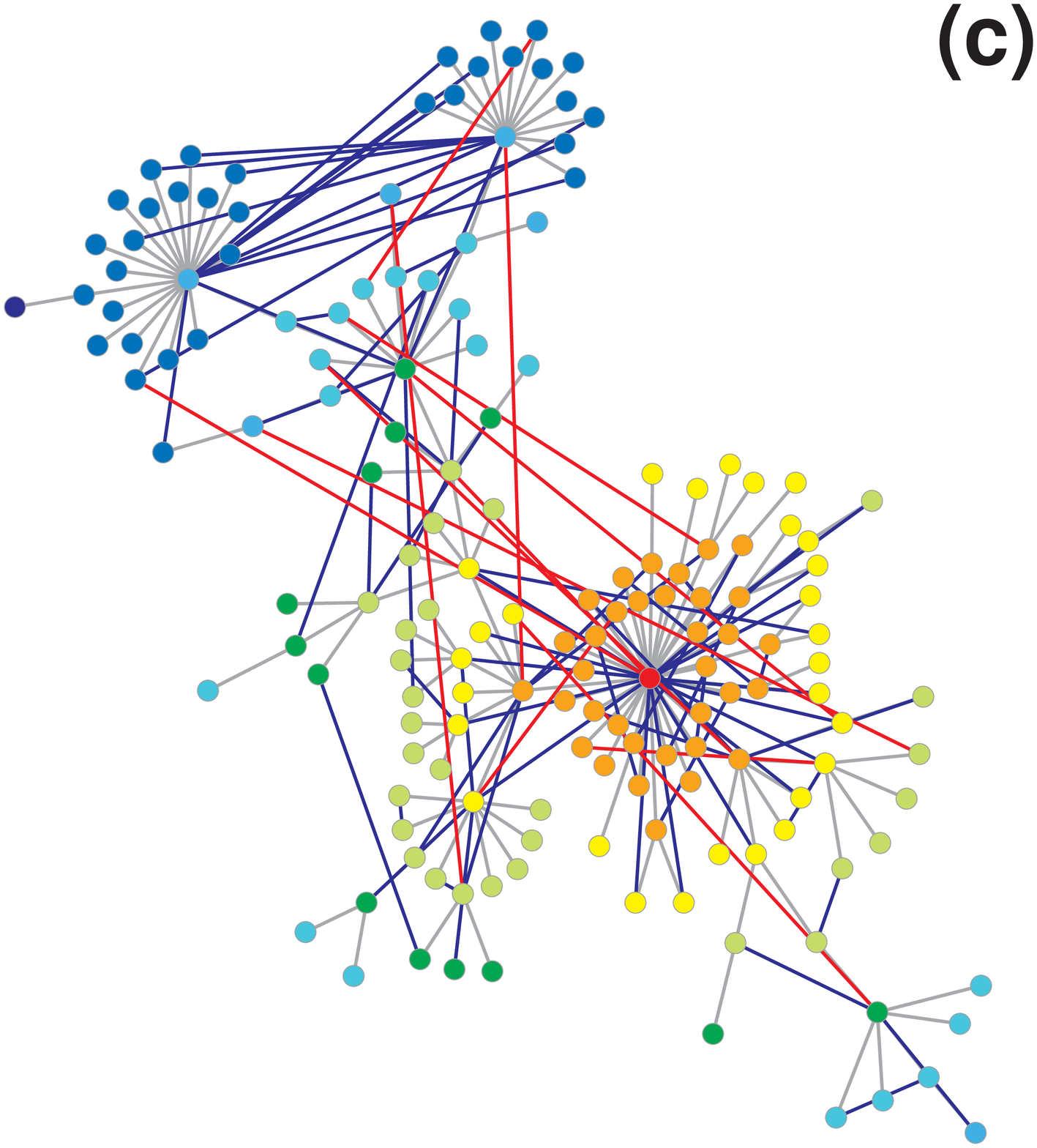}}
\end{minipage}
\end{minipage}
\begin{minipage}{18cm}
\begin{minipage}{0.280\linewidth}
{\epsfxsize=\linewidth \epsfbox{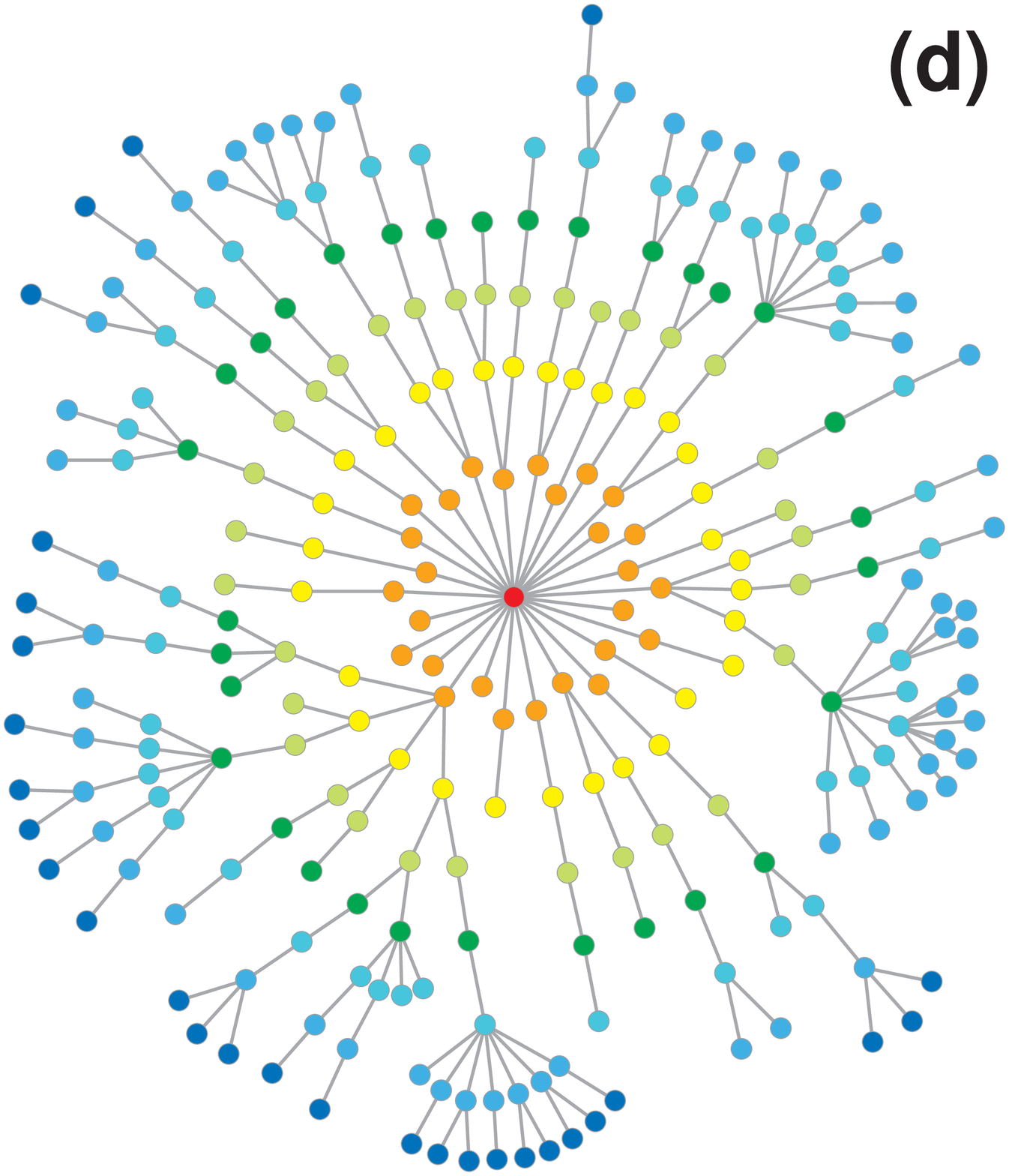}}
\end{minipage}
\begin{minipage}{0.280\linewidth}
{\epsfxsize=\linewidth \epsfbox{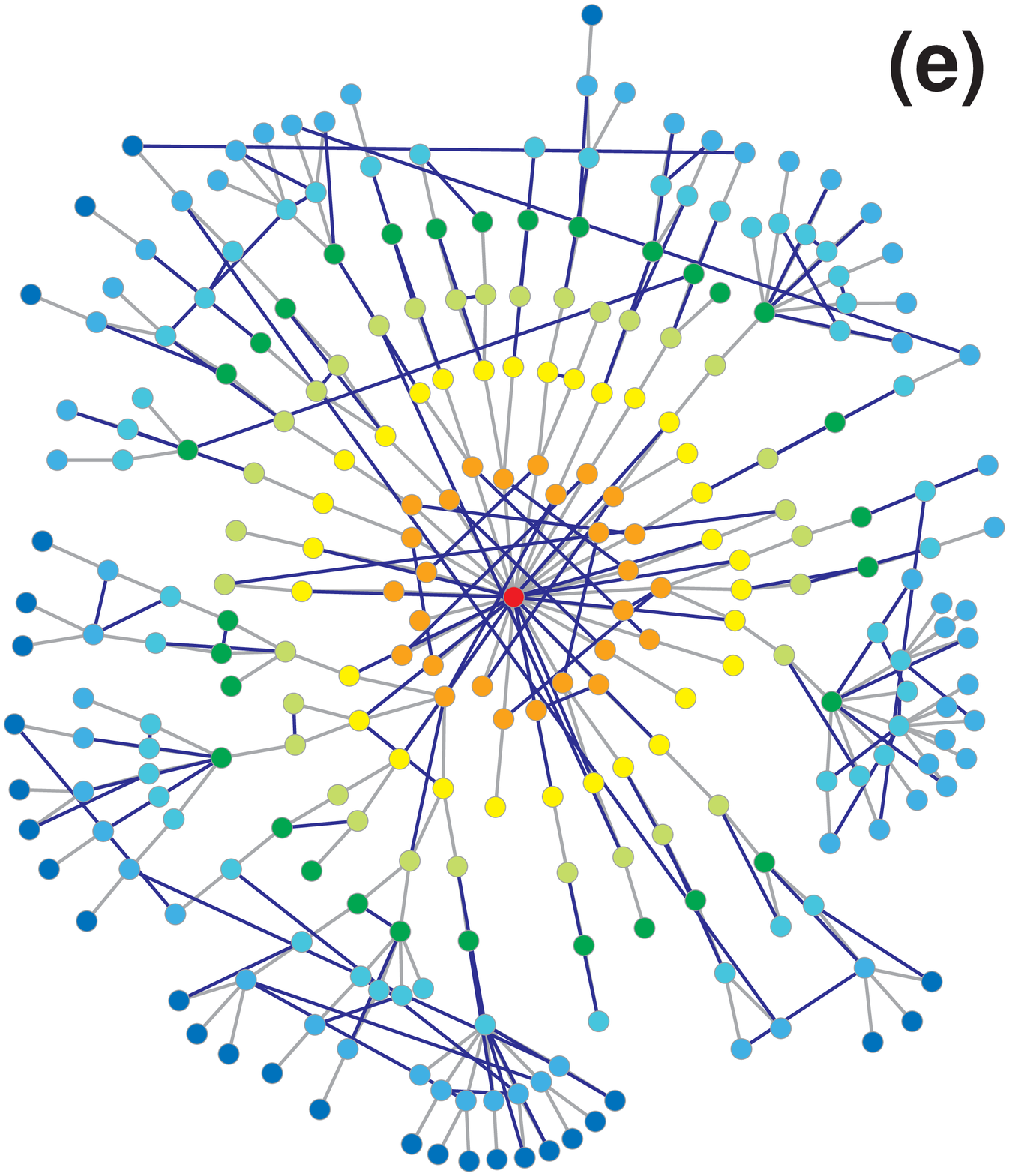}}
\end{minipage}
\begin{minipage}{0.280\linewidth}
{\epsfxsize=\linewidth \epsfbox{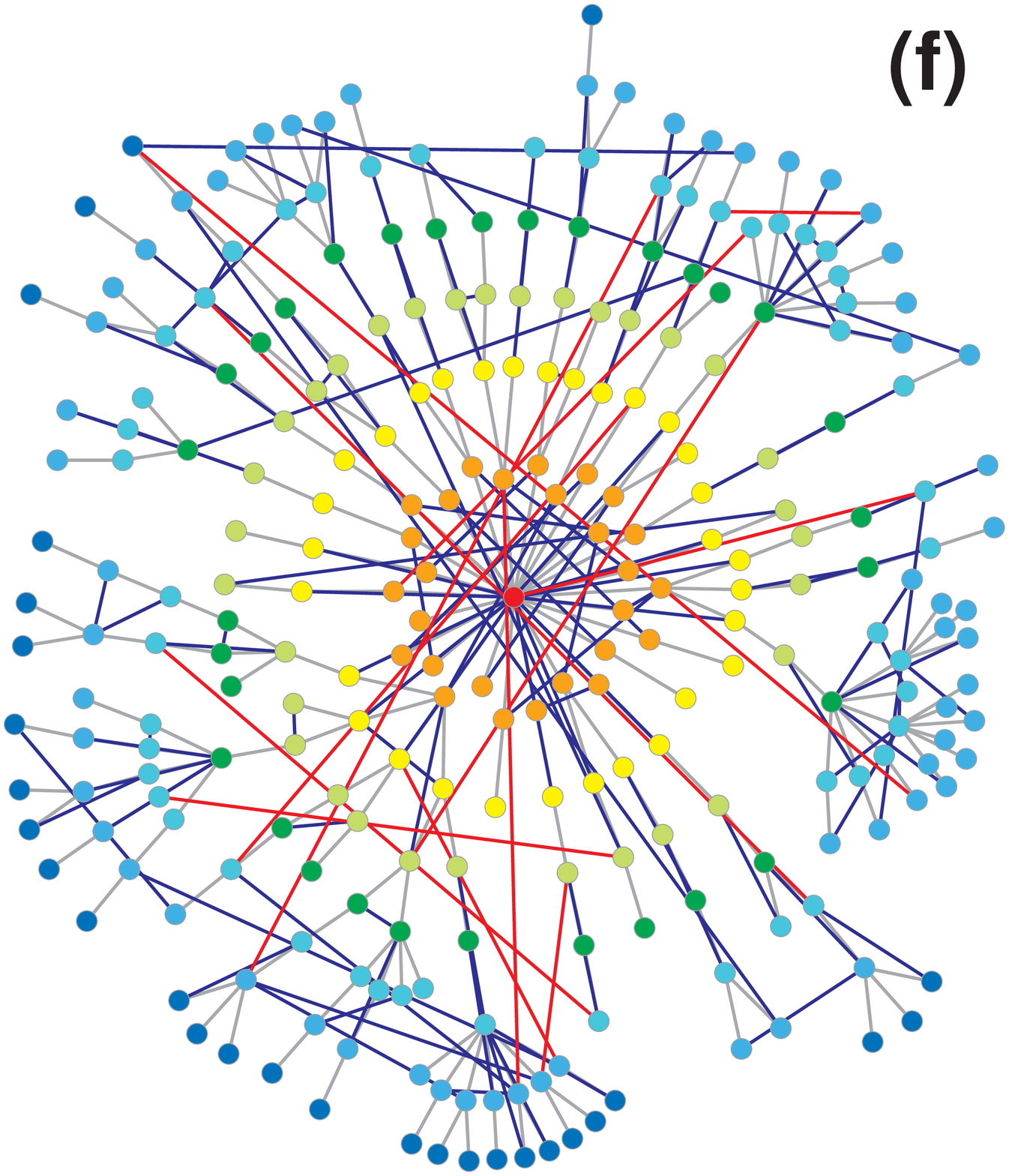}}
\end{minipage}
\end{minipage}
\caption{(Color online) Snapshots of the fractal network models.
(a) A critical branching tree with $\gamma=2.3$ and $N=164$
created in step (i). (b) A network created by adding local
shortcuts (blue) following the rule (ii) to the branching tree in
(a). Parameter $p=0.5$ is used. This network is still fractal. (c)
A network created by adding global shortcuts (red) following the
rule (iii) to the network in (b). Parameter $q=0.02$ is used. This
network is no longer fractal, but small-world. (d) A supercritical
branching tree with mean branching number $\langle n \rangle=2$,
which is fractal as well as small-world. (e) A dressed network to
network (d) by local shortcuts (blue) generated with $p=0.5$. The
network is fractal as well as small-world. (f) A network created
by adding global shortcuts (red) to the network in (e). $q=0.02$
is used. In (a)--(f), the colors of each vertex represent distinct
generations from the root.} \label{model}
\end{figure*}

Network configurations obtained by the model with $\gamma=2.3$ and
$N=164$ are shown in Fig.~\ref{model}. A critical branching tree
with $\langle n\rangle=1$ is shown in Fig.~\ref{model}(a); this
tree is dressed by local [Fig.~\ref{model}(b)] generated with
$p=0.5$ and $q=0$ and both local and global shortcuts
[Fig.~\ref{model}(c)] with $p=0.5$ and $q=0.02$. A supercritical
branching tree with $\langle n \rangle=2$ is shown in
Fig.~\ref{model}(d); this tree is dressed by both local and global
shortcuts generated with parameters $p=0.5$ and $q=0.02$ in
Figs.~\ref{model}(e) and (f), respectively.

\begin{figure}
\centerline{\epsfxsize=9cm \epsfbox{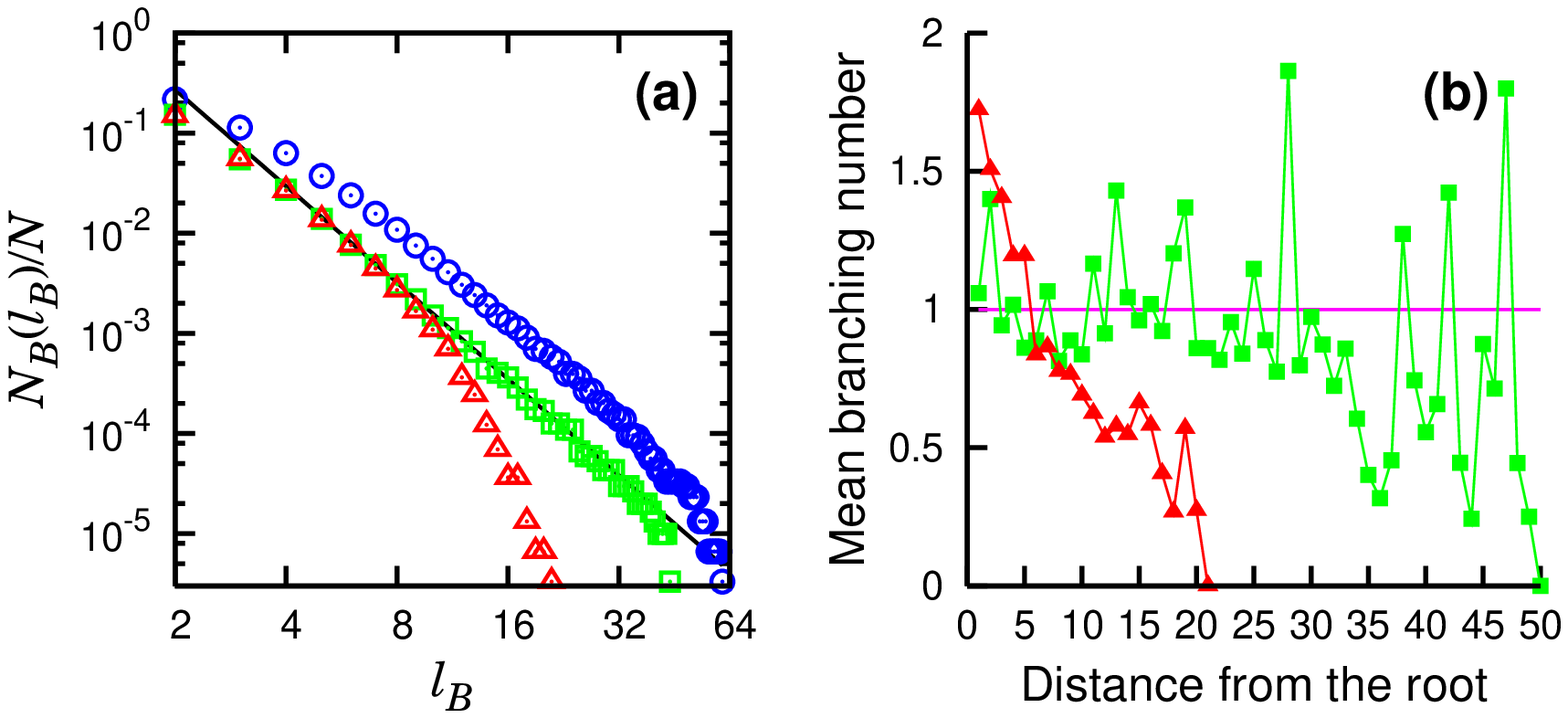}} \caption{(Color
online) Fractal scaling analysis (a) and mean branching number (b)
for the fractal models based on the critical branching tree with
only local shortcuts with $p=0.5$ and $q=0$
(\textcolor{green}{$\square$}), and with 1\% of global shortcuts
with $p=0.5$ and $q=0.01$ (\textcolor{red}{$\triangle$}). The bare
critical branching tree is represented by
(\textcolor{blue}{$\bigcirc$}). The solid line in (a) is guideline
with a slope of --3.2. The measured degree exponent is $\gamma
\approx 2.4$ and system size is $N=3\times 10^5$.}
\label{fig7_model1}
\end{figure}

We examine the fractal scaling in the network model and the MBN
for its skeleton. In the case of a network generated from a
critical branching tree (with $\gamma=2.3$ and $N\approx
3\times10^5$) and dressed only by local shortcuts (with $p=0.5$
and $q=0$), 76\% of all edges of the original branching tree are
maintained in the skeleton. The branching tree and the dressed
network exhibit fractal scalings with the same fractal exponent
$d_B\approx3.2$ [Fig.~\ref{fig7_model1}(a)]. This value appears to
differ from the theoretical value $\approx 4.3$ estimated from the
formula (\ref{z}). However, we notice that the measured value of
the degree exponent of the dressed network is rather close to
$\gamma=2.4$, although the branching tree is generated with
parameter $\gamma=2.3$. Thus, the expected value is $d_B=3.5$.
Therefore, the numerical deviation can be explained. The MBN of
the skeleton of the dressed network displays a plateau around 1
[Fig.~\ref{fig7_model1}(b)]. Moreover, when we introduce 1\% of
global shortcuts ($p=0.5$ and $q=0.01$) to the critical branching
tree, the box number $N_B(\ell_B)$ decays faster than any power
law for large values of $\ell_B$ [Fig.~\ref{fig7_model1}(a)]; thus
fractality is lost. Accordingly, in this case, the MBN of the
skeleton decays to zero without a plateau
[Fig.~\ref{fig7_model1}(b)]. The critical value $q_c$ above which
the network becomes non-fractal depends on the degree exponent
$\gamma$, system size $N$, and number of shortcuts $p$. A more
detailed analysis on this crossover behavior will be presented
elsewhere \cite{jskim2006}.

\begin{figure}
\centerline{\epsfxsize=9cm \epsfbox{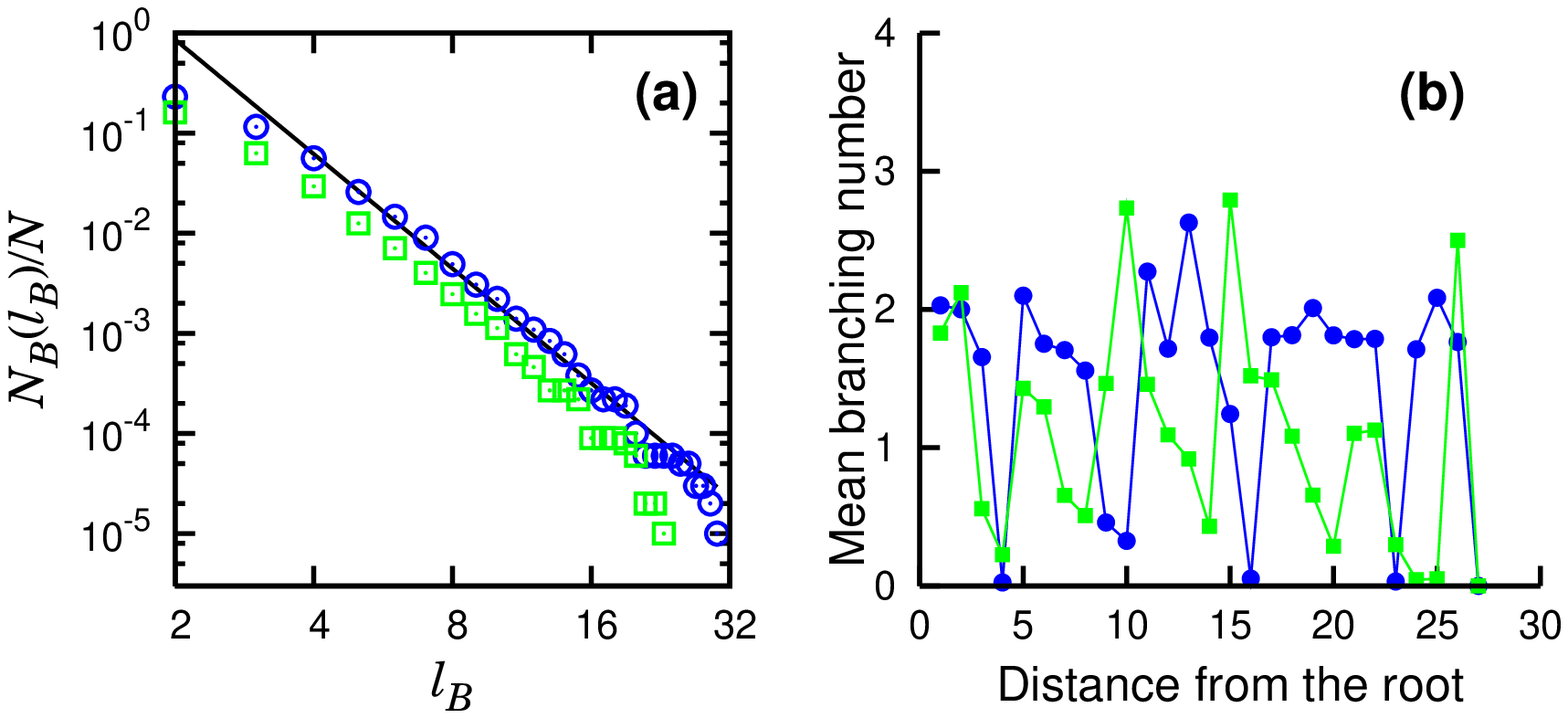}}
\centerline{\epsfxsize=9cm \epsfbox{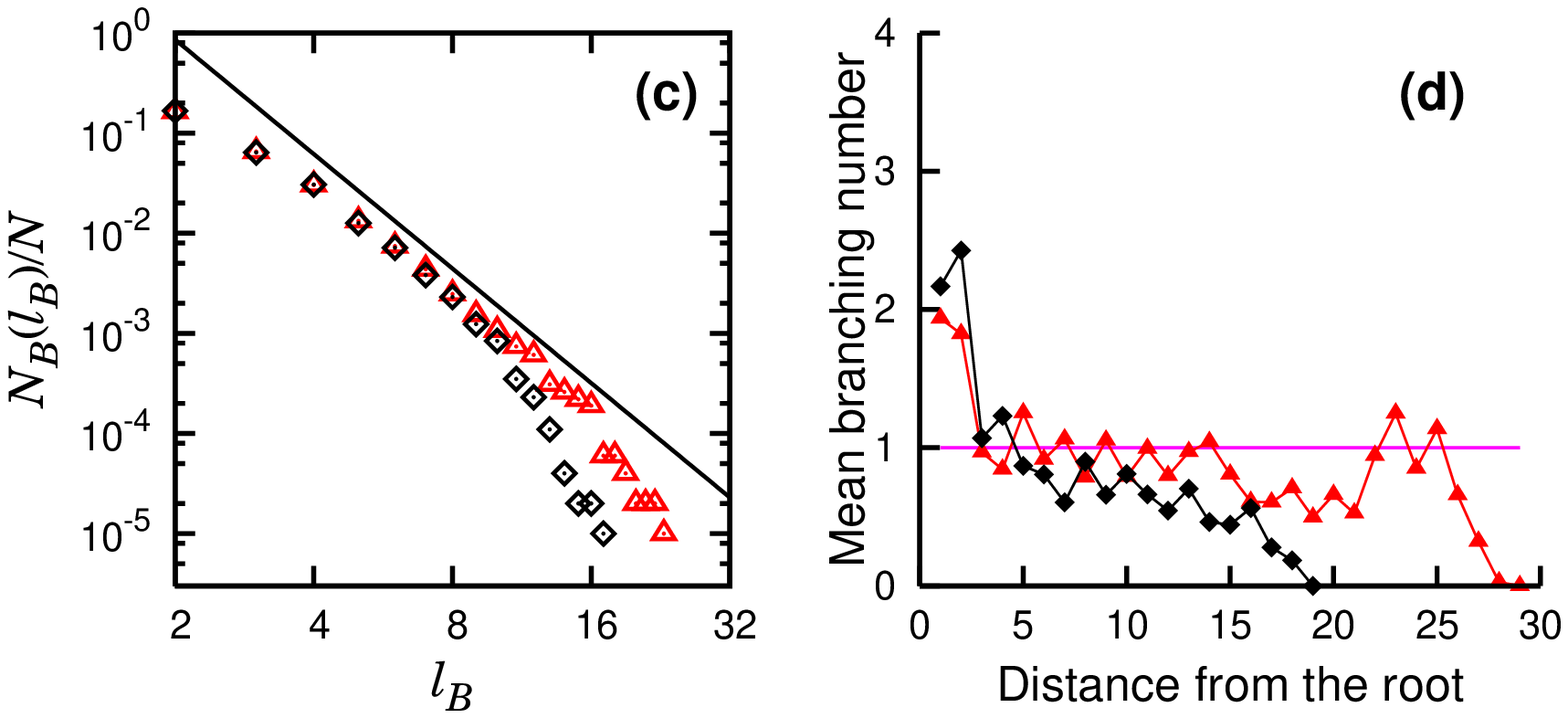}} \caption{(Color
online) Fractal scaling analysis (a) and (c), and mean branching
number (b) and (d) for the fractal models generated from a {\it
supercritical} branching tree with $\langle n \rangle=2$, dressed
by shortcuts. The data are for the bare supercritical tree
(\textcolor{blue}{$\circ$}) and dressed networks with $p=0.5$ and
$q=0$ (\textcolor{green}{$\square$}), $p=0.5$ and $q=0.01$
(\textcolor{red}{$\triangle$}), and $p=0$ and $q=0.05$
($\diamond$). The solid lines in (a) and (c) are guidelines with
slopes of --4.2 each. The degree exponent is $\gamma=2.3$ and
system size is $N=1\times 10^5$. }\label{fig8_model2}
\end{figure}

The same analysis is performed for the model based on the
supercritical branching tree with $\langle n \rangle=2$
[Fig.~\ref{fig8_model2}]. This tree with $\gamma=2.3$ displays a
power-law fractal scaling with fractal exponent $d_B\approx4.2$
[Fig.~\ref{fig8_model2}(a)]; however, its MBN fluctuates heavily
on and off about the expected value $\bar{n}=2$, while exhibiting
persistent branching [Fig.~\ref{fig8_model2}(b)]. Such a highly
fluctuating MBN is similar to that observed in the skeleton of the
WWW or the metabolic network
[Figs.~\ref{fig3}(a$^{\prime}$)--(b$^{\prime}$)]. When dressed
only by local shortcuts ($p=0.5$ and $q=0$), the dressed network
still exhibits a power-law fractal scaling
[Fig.~\ref{fig8_model2}(a)]. The MBN of its skeleton still exhibit
large fluctuations [Fig.~\ref{fig8_model2}(b)]; however, its mean
$\bar n$ decreases from 2. With 1\% of global shortcuts
($q=0.01$), the fractal scaling exhibits a power-law behavior but
with an exponential cutoff [Fig.~\ref{fig8_model2}(c)].
Interestingly, the MBN of its skeleton displays a plateau located
around 1 with reduced fluctuations [Fig.~\ref{fig8_model2}(d)].
When we further increase the number of global shortcuts to 2\%
($q=0.02$), the MBN of the skeleton decays without a plateau, and
the network is no longer a fractal
[Figs.~\ref{fig8_model2}(c)--(d)].

The fractal network model studied here is a generalization of the
previous model \cite{goh2006} generated by including the
supercritical branching case. Using this model, we can reproduce
the highly fluctuating behavior in the MBN observed in the WWW as
well as the fractal scaling. We will also show that this
generalization of the supercritical branching facilitates a better
understanding of how the SW and fractal scaling coexist and do not
contradict each other in such systems. In the following, we
investigate the properties of the fractal network using the
fractal network model as well as the real-world fractal networks.

\section{Small-worldness and Box mass distribution}

In this section, we study the average box mass $\langle
M(\ell)\rangle$ as a function of box size $\ell$. The box mass is
measured using two methods, the cluster-growing method and the
box-covering method \cite{ss}. In the cluster-growing method, the
box mass is defined as the number of vertices at a distance not
greater than $\ell_C$ from a given vertex. Note that in the
cluster-growing method, a vertex can be counted by more than one
box, whereas in the box-covering method, it is counted only once.
The cluster-growing method provides information on the SW of the
network. The average box mass for the {\em critical} branching
tree grows with distance $\ell_C$, according to a power law with
the exponent $d_B$,
\begin{equation} \langle
M_C(\ell_C)\rangle \sim \ell_C^{d_B} \label{fractal_mass},
\end{equation}
where $d_B$ is defined in Eq.~(\ref{z}); this implies that the
critical branching tree is a fractal. For the supercritical
branching tree, the average mass grows exponentially with increase
in distance $\ell_C$,
\begin{equation} \langle
M_C(\ell_C)\rangle \sim \langle n\rangle^{\ell_C}. \label{sw_mass}
\end{equation}
This relation is equivalent to Eq.~(\ref{sw}), thereby suggesting
that the supercritical branching tree is a small-world network. On
the other hand, the average box mass in the box-covering method is
determined by the fractality. Since both the critical and
supercritical branching trees are fractals, their box mass
increases according to a power law. Thus, the analysis of the
average box mass in the two methods will provide an insight into
the interplay between SW and fractality.

We also consider the distribution of box masses $P_m(M)$ for each
method. It is known \cite{ss} that in the cluster-growing method,
the box mass distribution exhibits a peak at a characteristic
mass, while it exhibits a fat tail without a peak in the
box-covering method. However, the origin of the power-law behavior
of the box mass distribution for the fractal networks has not been
understood clearly. Here, we present a detailed analysis of the
box mass distribution, showing that the exponent of the power-law
behavior depends on the lateral size of the box. When the lateral
size of the box is large, the exponent of the box-mass
distribution can be understood from the perspective of branching
dynamics.

\begin{figure}
\centerline{\epsfxsize=9cm \epsfbox{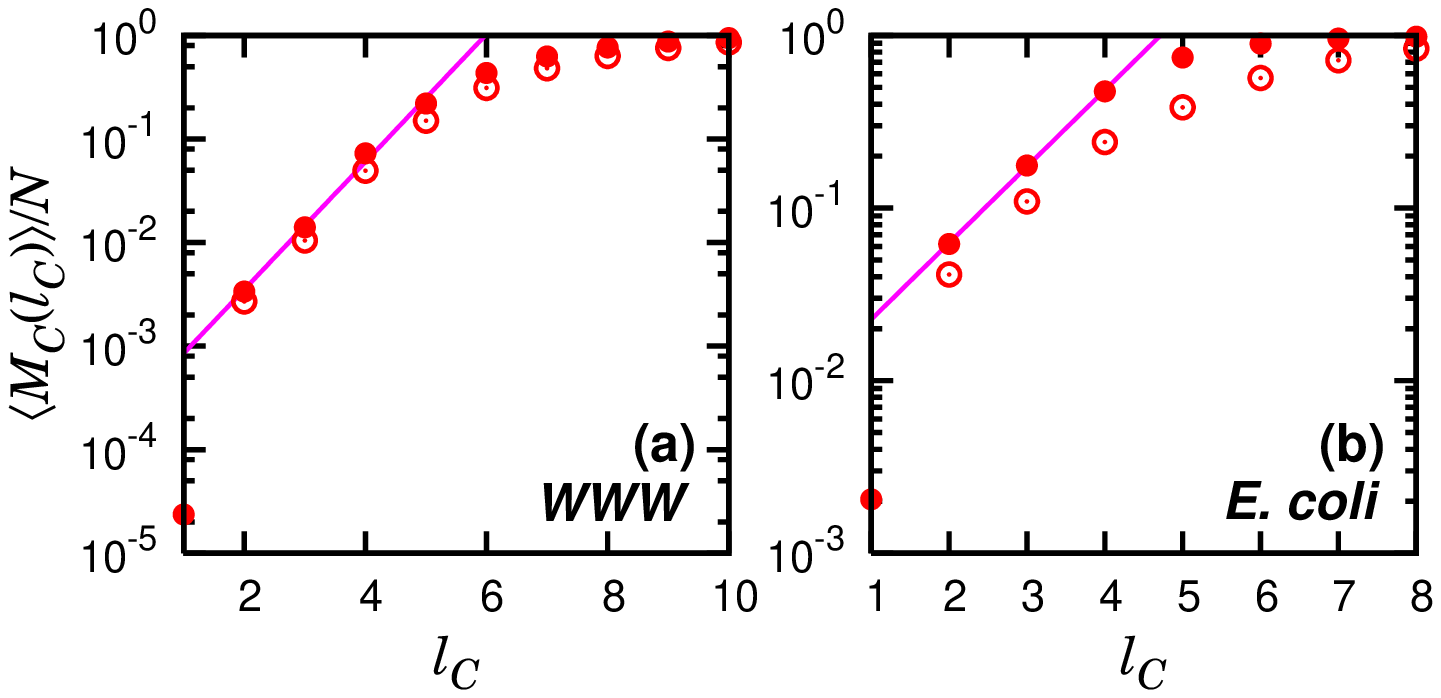}}
\centerline{\epsfxsize=9cm \epsfbox{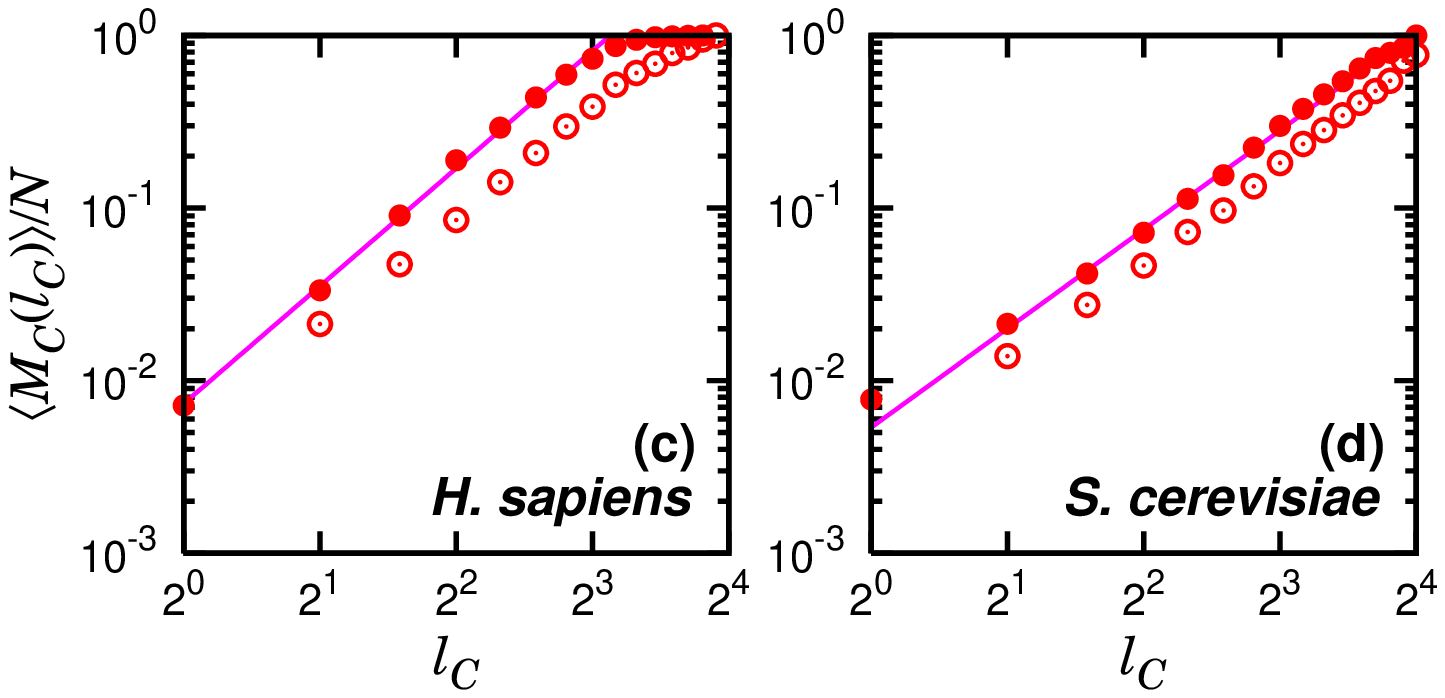}} \caption{Average
box mass $\langle M_C (\ell_C) \rangle$ in the cluster-growing
method divided by the total number of vertices $N$, as a function
of the distance $\ell_C$. Plots of (a) and (b) are drawn on a
semi-logarithmic scale and those of (c) and (d) on a
double-logarithmic scale, respectively. Filled and open symbols
represent the original network and the skeleton of each network,
respectively. The solid lines for reference in (c) and (d) have
slopes of 1.9 and 2.3, respectively.} \label{avemass_cluster}
\end{figure}

\subsection{Real-world fractal networks}

We first examine the average box mass $\langle M_C(\ell_C)
\rangle$ in the cluster-growing method for the original network
and the skeleton of each real-world fractal network. For the WWW,
we find that both the original network and the skeleton exhibit an
exponential increase in average box mass with distance $\ell_C$
[Fig.~\ref{avemass_cluster}(a)]. Thus, the WWW is a small-world
network and the skeleton of the WWW is a {\em supercritical}
branching tree. For the metabolic network, while the original
network is small-world with Eq.~(\ref{sw_mass}), its skeleton
appears to follow a power law, Eq.~(\ref{fractal_mass})
[Fig.~\ref{avemass_cluster}(b)]. Thus, the skeleton of the
metabolic network is better described by a critical branching
tree, although the original network is small-world. The difference
between the metabolic network and its skeleton probably originates
from the presence of core subnetworks in the metabolic network,
wherein the vertices are tightly interwoven through multiple
pathways but are simplified into a tree in the skeleton
\cite{pnas,almaas,cmghim}. On the other hand, for the protein
interaction networks, both the original networks and the skeletons
behave according to a power-law form of Eq.~(\ref{fractal_mass}).
Therefore, the protein interaction networks are not likely to be
small-world and their skeletons can be regarded as critical
branching trees [Figs.~\ref{avemass_cluster}(c)--(d)].

\begin{figure}
\centerline{\epsfxsize=9cm \epsfbox{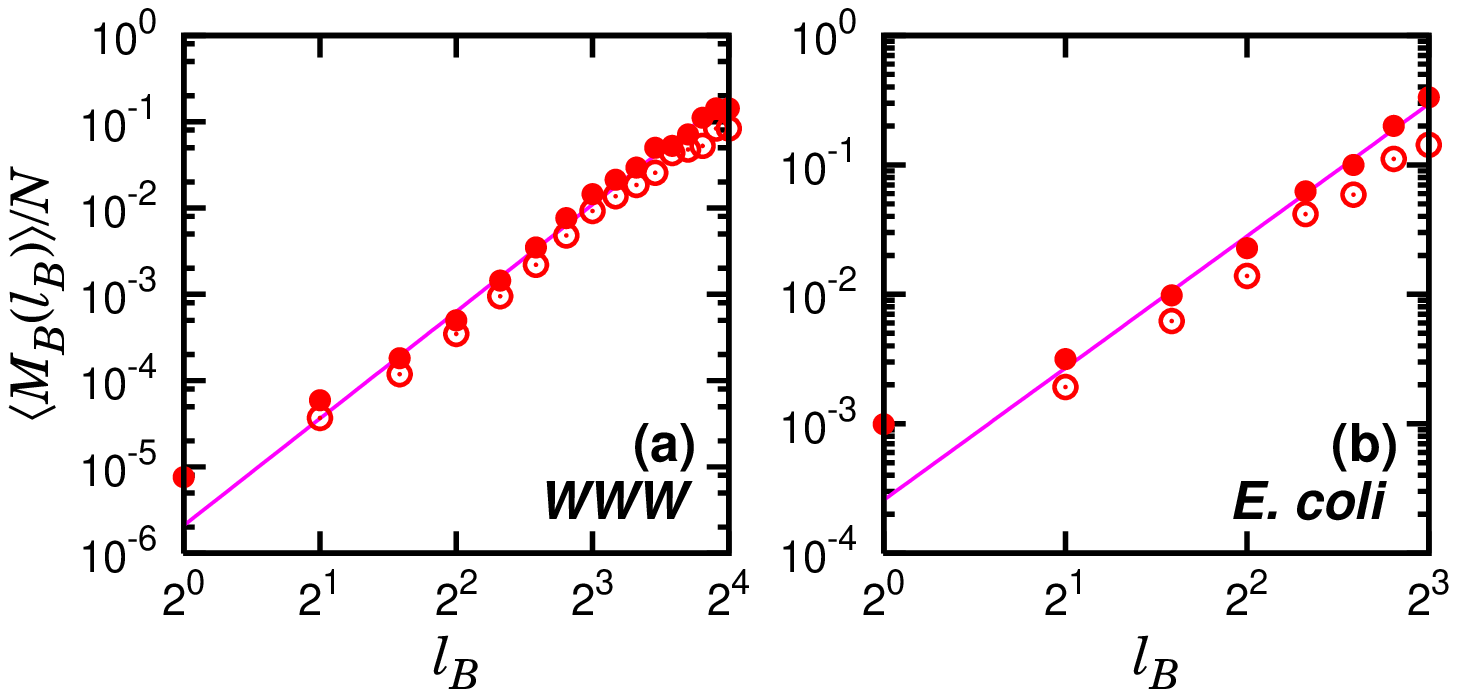}}
\centerline{\epsfxsize=9cm \epsfbox{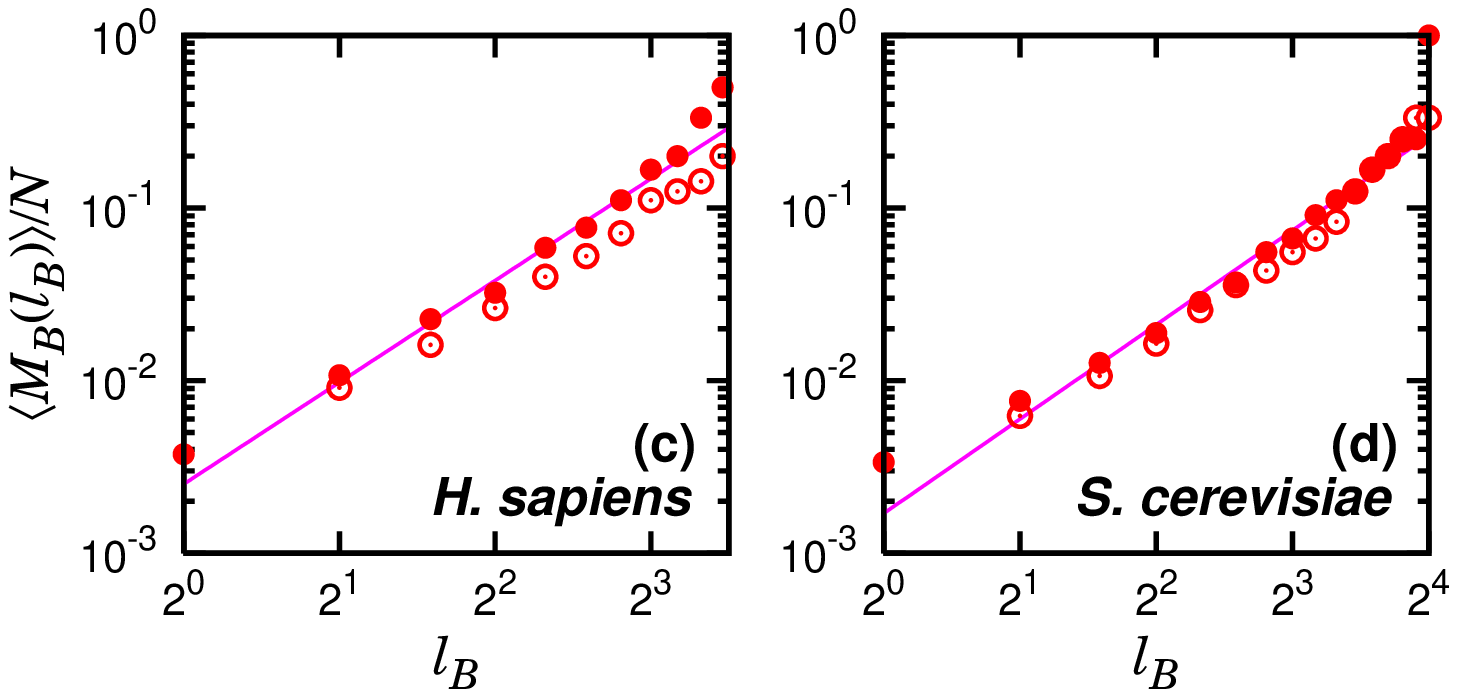}} \caption{Average
box mass $\langle M_B (\ell_B) \rangle$ divided by the total
number of vertices $N$, as a function of box size $\ell_B$ in the
box-covering method. Filled and open symbols represent the
original network and the skeleton of each network, respectively.
The solid lines (drawn for reference) have slopes of 4.1, 3.4,
2.0, and 1.8 for (a), (b), (c), and (d), respectively.}
\label{avemass_cluster2}
\end{figure}

\begin{table}[b]
\caption{Behavior of the average box mass of the fractal networks
and their skeletons in the cluster-growing and box-covering
methods.}
\begin{ruledtabular}
\begin{tabular}{lll}
& Cluster-growing & Box-covering \\
& method & method
\\
\hline World-wide web & Exponential & Power law \\
World-wide web (skeleton) & Exponential & Power law \\
Metabolic network & Exponential & Power law \\
Metabolic network (skeleton) & Power law & Power law \\
PIN of {\it H. sapiens} & Power law & Power law \\
PIN of {\it H. sapiens} (skeleton) & Power law & Power law \\
PIN of {\it S. cerevisiae} & Power law & Power law \\
PIN of {\it S. cerevisiae} (skeleton) & Power law & Power law
\end{tabular}
\end{ruledtabular}
\end{table}

Next, we study the average box mass $\langle M_B(\ell_B) \rangle$
in the box-covering method. For the fractal networks and their
skeletons, the average mass $\langle M_B(\ell_B)\rangle$ increases
according to a power law with respect to box size $\ell_B$ of
Eq.~(\ref{boxcovering}), regardless of whether it is critical or
supercritical in the cluster-growing method as shown in
Fig.~\ref{avemass_cluster2}. The fractal dimensions measured using
the formula (\ref{boxcovering}) are $d_B=4.1, 3.4, 2.0$, and 1.8
for the WWW (a), metabolic network (b), protein interaction
network of {\it H. sapiens} (c), and {\it S. cerevisiae} (d),
respectively. These values are comparable to the ones obtained
from the fractal scaling (\ref{fractal}), which are $d_B=4.1$,
3.5, 2.3, and 2.1 for (a), (b), (c), and (d), respectively. The
results of the average box mass for the real-world fractal
networks are summarized in Table II. Non-fractal networks exhibit
the exponential relationship
\begin{equation} \langle
M_B(\ell_B)\rangle \sim \exp(\ell_B/\ell_0)\label{boxcoveringsw}
\end{equation} with a constant $\ell_0$.

\begin{figure}
\centerline{\epsfxsize=6cm \epsfbox{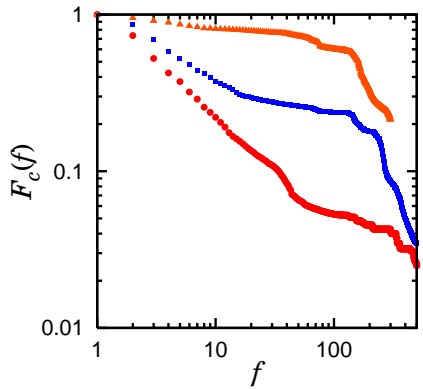}} \caption{(Color
online) Cumulative fraction $F_c(f)$ of the vertices counted $f$
times in the cluster-growing algorithm. $F_c(f)$ follows a power
law in the small $f$ region, where the slope depends on box size
$\ell_C$. However, for large values of $f$, the data largely
deviate from the value extrapolated from the power-law behavior.
Data are presented for $\ell_C=2$ (\textcolor{red}{$\bullet$}),
$\ell_C=3$ (\textcolor{blue}{$\blacksquare$}), and $\ell_C=5$
(\textcolor{orange}{$\blacktriangle$}).} \label{histogram}
\end{figure}

The different behaviors of the average mass in the two methods,
the cluster-growing and box-covering methods, originates from
whether overlap between the boxes is allowed. Thus, studying the
extent of overlap of the boxes during the tiling can provide
important information. In this regard, we measure the cumulative
fraction $F_c(f)$ of vertices counted $f$ times or more in the
cluster-growing method for the WWW in Fig.~\ref{histogram}. The
cumulative fraction $F_c(f)$ is likely to follow a power law for
small $f$, thereby indicating that the overlaps occur in a
non-negligible frequency even for a small distance $\ell_C$. The
associated exponent decreases with increase in box size $\ell_C$
as the chances of overlaps increase. However, for large values of
$f$, the large fraction of vertices counted exceed the frequency
extrapolated from the power-law behavior. As opposed to a bounded
distribution such as a Poisson-type distribution, the broad
distribution of $f$ implies that a significant fraction of
vertices are counted more than once in the cluster-growing method.
Such multiple counting due to overlap is excluded in the
box-covering method. Due to this exclusion effect, the mass of a
box in the box-covering method is significantly lower than that in
the cluster-growing method.

\begin{figure}
\centerline{\epsfxsize=9.5cm \epsfbox{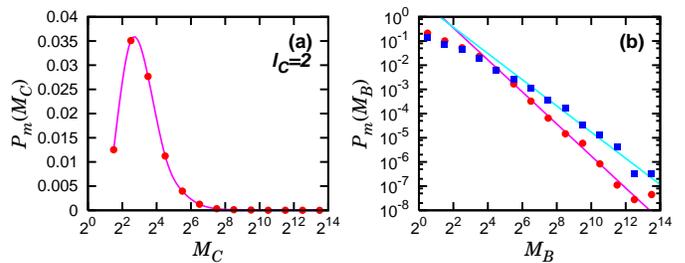}} \caption{(Color
online) Box mass distribution in the cluster-growing method (a)
and box-covering method (b) for the WWW. Data in (a) are for
$\ell_C=2$ and those in (b) are for $\ell_B=2$
(\textcolor{red}{$\bullet$}) and $\ell_B=5$
(\textcolor{blue}{$\blacksquare$}). The solid lines are guidelines
with slopes of -2.2 and -1.8, respectively.} \label{boxmass_dist}
\end{figure}

We study the box-mass distributions in the two methods. As shown
in \cite{ss}, for the WWW, the box-mass distribution in the
cluster-growing method exhibits a clear peak
[Fig.~\ref{boxmass_dist}(a)]; on the other hand, in the
box-covering method, it exhibits a fat tail, following an
asymptotic power law,
\begin{equation} P_m(M_B)\sim M_B^{-\eta} \label{PmMB} \end{equation}
[Fig.~\ref{boxmass_dist}(b)]. We find that the exponent $\eta$
depends on the box size $\ell_B$. For small $\ell_B=1$ or 2, it is
found that $\eta$ is equal to $\gamma$; however, as $\ell_B$
increases, $\eta$ approaches the exponent $\tau$ of the
cluster-size distribution (\ref{tau}). This can be understood as
follows. For small values of $\ell_B$, the branching has not
proceeded sufficiently to exhibit asymptotic behavior; thus, the
box mass will simply scale with the degree of the seed vertex,
which is selected randomly, yielding $\eta=\gamma$. This is most
evident for $\ell_B=1$. On the other hand, as $\ell_B$ increases,
the box grows and its size governs the scaling. The growth of the
box can be approximated by the SF branching tree with the exponent
$\gamma$, the size distribution of which follows a power law with
the exponent $\tau$ given by Eq.~(\ref{tau}), yielding $\eta=\tau$
for large $\ell_B$. The numerical estimates of $\eta$ obtained
from the WWW are in reasonable agreement with the prediction as
$\eta=\gamma\approx 2.3$ for $\ell_B=2$ and $\eta\approx 1.8$ for
$\ell_B=5$ [Fig.12].

\begin{figure}
\centerline{\epsfxsize=9cm \epsfbox{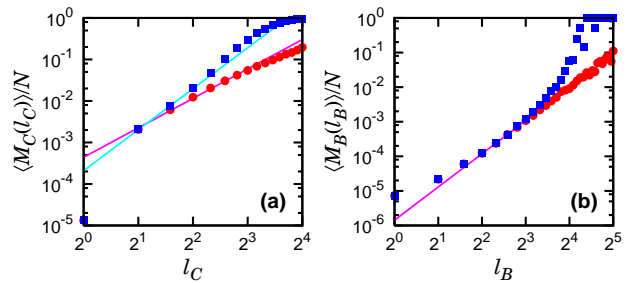}} \caption{Average
box mass versus box lateral size in the cluster-growing method (a)
and the box-covering method (b) for the fractal network model
constructed from a {\em critical} branching tree, dressed by
shortcuts with $p=0.5$ and $q=0$ (\textcolor{red}{$\bullet$}) and
$p=0.5$ and $q=0.01$ (\textcolor{blue}{$\blacksquare$}). Degree
exponent is $\gamma=2.3$ and system size is $N=3\times 10^5$.
Solid lines in (a) have slopes of 3.3 and 2.4, respectively, and
the solid line in (b) has a slope of 3.3.} \label{avemass1_model}
\end{figure}
\begin{figure}
\centerline{\epsfxsize=9cm \epsfbox{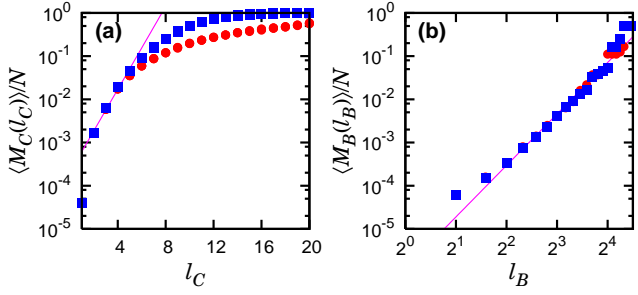}} \caption{(Color
online) Average box mass as a function of box size in the
cluster-growing method (a) and the box-covering method (b) for the
model network constructed from a {\it supercritical} branching
tree, dressed by shortcuts with $p=0.5$ and $q=0$
(\textcolor{red}{$\bullet$}) and $p=0.5$ and $q=0.01$
(\textcolor{blue}{$\blacksquare$}). Degree exponent is
$\gamma=2.3$, and system size $N=1\times 10^5$. The solid line in
(b) has a slope of 4.0, which is drawn for guidance.}
\label{avemass2_model}
\end{figure}

\subsection{Fractal network model}

Here, we perform a similar analysis of the average box mass and
the box-mass distribution for the fractal network model introduced
in Sec.~V. We first consider a network model based on a {\em
critical} branching tree with mean branching number $\langle n
\rangle=1$. For simplicity, we fix the parameters to be
$\gamma=2.3$, $N=3\times10^5$, and $p=0.5$, while varying
parameter $q$. When $q$ is sufficiently small, i.e., $q\le 0.001$,
the model network exhibits a power-law scaling both in the
cluster-growing method Eq.~(\ref{fractal_mass}) and the
box-covering method Eq.(\ref{boxcovering}). The network thus
remains as a fractal. However, for larger values of $q$ like
$0.01$, the fractal scaling breaks down and the average box mass
increases exponentially as Eq.~(\ref{sw_mass}) in both methods
[Fig.~\ref{avemass1_model}], i.e., the network becomes
small-world.

The behavior of the average box mass of the fractal model network
based on the supercritical tree is interesting. Once a
supercritical tree is generated, the model network is dressed by
local shortcuts with $p=0.5$. Then it simultaneously exhibits both
an exponential increase in box mass in the cluster-growing method
and a power-law increase in the box-covering method, as observed
in the WWW. This coexistence persists when we introduce global
shortcuts up to $q=0.01$ [Fig.~\ref{avemass2_model}]. If we
further increase $q$, the average box mass increases exponentially
in both methods as the network loses fractality. Thus, the model
network with supercritical branching tree and an appropriate
number of local shortcuts can reproduce the small-world property
of the average box-mass function as well as the fractality.

\begin{figure}
\centerline{\epsfxsize=10cm \epsfbox{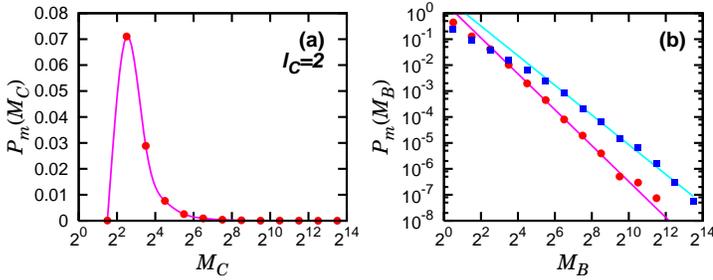}} \caption{(Color
online) Box-mass distribution in the cluster-growing method (a)
and box-covering method (b) for the fractal model network grown
from a {\it critical} branching tree with $\gamma=2.3$ and dressed
by shortcuts generated with $p=0.5$ and $q=0.001$. The data in (b)
are for $\ell_B=2$ (\textcolor{red}{$\bullet$}) and $\ell_B=5$
(\textcolor{blue}{$\blacksquare$}). Their slopes are --2.3 and
--1.8, respectively. The system size is $N\approx 3\times 10^5$.}
\label{boxmass_dist_model}
\end{figure}
\begin{figure}
\centerline{\epsfxsize=10cm \epsfbox{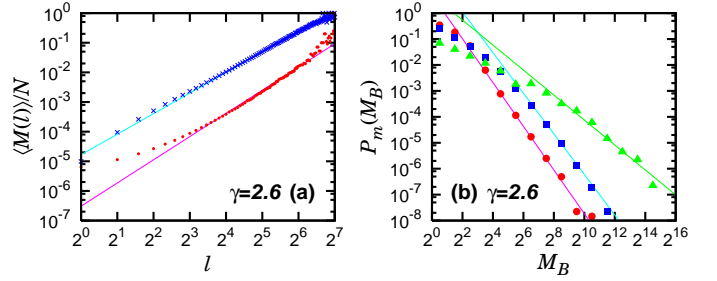}} \caption{(Color
online) (a) Average box mass versus box size in the
cluster-growing (\textcolor{blue}{$\times$}) and box-covering
(\textcolor{red}{$\blacksquare$}) methods for a bare critical
branching tree with $\gamma=2.6$. Solid lines, drawn for guidance,
have slopes of 2.3 (\textcolor{blue}{$\times$}) and 2.6
(\textcolor{red}{$\blacksquare$}), respectively.  (b) Box-mass
distribution for the bare tree of (a). Solid guidelines have
slopes of --2.8 for $\ell_B=2$ (\textcolor{red}{$\bullet$}), --2.6
for $\ell_B=5$ (\textcolor{blue}{$\blacksquare$}), and --1.6 for
$\ell_B=32$ (\textcolor{green}{$\blacktriangle$}). }
\label{box_mass34}
\end{figure}

\begin{figure}
\centerline{\epsfxsize=10cm \epsfbox{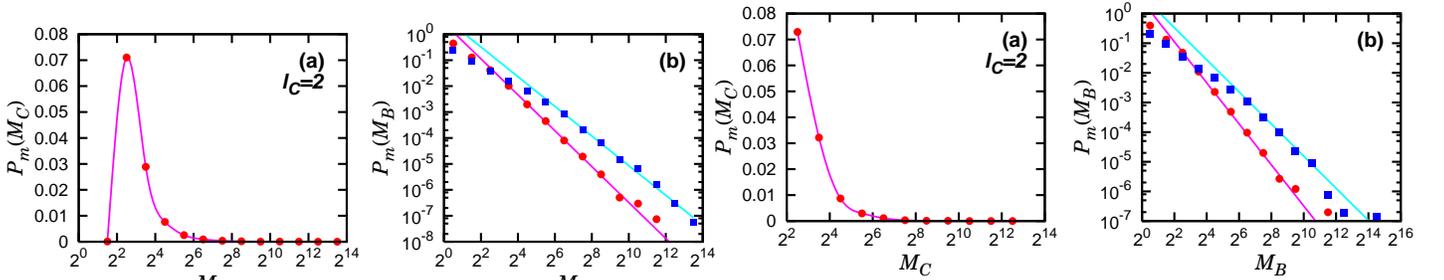}} \caption{(Color
online) Box-mass distribution in the cluster-growing method (a)
and box-covering method (b) for the fractal model network grown
from a supercritical branching tree with $\langle n \rangle=2$ and
$\gamma=2.3$, which is dressed by shortcuts with the parameters
$p=0.5$ and $q=0$. The box size in (b) is $\ell_M=2$
(\textcolor{red}{$\bullet$}) and $\ell_M=5$
(\textcolor{blue}{$\blacksquare$}). Solid lines in (b) have slopes
--2.3 and --1.8, drawn for guidance. The system size is $N\approx
1\times 10^5$.} \label{boxmass_dist_model2}
\end{figure}

Next, we study the box-mass distribution for the fractal network
model. We restart the analysis with the model network based on the
critical branching tree with parameters $\gamma=2.3$,
$N=3\times10^5$, and $p=0.5$. As with the real-world fractal
networks like the WWW, the box-mass distribution for the model
network exhibits a peak at a finite mass in the cluster-growing
method. The box-mass distribution in the box-covering method
follows an asymptotic power law with exponent $\eta$. As observed
for the WWW, we observe that the exponent $\eta$ depends on the
box size. For $q=0.001$, it is found that $\eta\approx2.3$ for
small $\ell_B=2$, and $\eta\approx1.8$ for large $\ell_B=5$
[Fig.~\ref{boxmass_dist_model}]. The latter value $\eta\approx
1.8$ is in agreement with $\tau=\gamma/(\gamma-1) \approx 1.8$
from Eq.~(\ref{tau}). Such $\ell_B$-dependent behavior of the box
mass distribution can also be observed for another value of
$\gamma$, for example, $\gamma=2.6$ [Fig.~\ref{box_mass34}]. In
such cases, the behavior $\eta=\tau$ appears for large values of
$\ell_B$, for example, $\ell_B=32$ for $\gamma=2.6$.

Next, when the model network is constructed based on a
supercritical branching tree (with $\gamma=2.3$ and $\langle n
\rangle=2$) and is dressed by shortcuts (with $p=0.5$ and
$q=0.001$), $\eta \approx 2.3$ is measured for $\ell_B=2$,
however, $\eta\approx 1.8$ for $\ell_B=5$
[Fig.~\ref{boxmass_dist_model2}]. The obtained value $\eta\approx
1.8$ is again in agreement with
the expected value $\tau\approx 1.8$ for $\gamma=2.3$.\\

\section{perimeter of a box}

The boundary of a fractal object is an important physical quantity
and is considered to be another fractal object. For example, the
area of the spin domain of the Ising model at critical
temperature, which corresponds to magnetization, is a fractal
object, and the interface length of the spin domain is another
fractal object, corresponding to the singular part of internal
energy~\cite{nijs}. Moreover, a percolation cluster is a fractal
object and its outer boundary, referred to as ``hull," is also a
fractal~\cite{leath,voss}.

We define the perimeter $H_{\alpha}$ of a given box $\alpha$ as
the number of edges connected on one end to the vertices within
the box $\alpha$ and on the other end to the vertices in other
boxes. The perimeter $H_{\alpha}$ is examined as a function of the
box mass $M_{B, \alpha}$ of the box $\alpha$. Then, we can define
the average perimeter $\langle H(M_B) \rangle$ over the boxes with
box mass $M_B$. We find that the following power-law relationship
exists, \be \langle H(M_B) \rangle \sim M_B^{d_H/d_B}.
\label{kMBalpha}\ee The new exponent $d_H$ (the hull exponent for
the fractal network) is analogous to the one used in the
percolation theory~\cite{leath,voss}.

\begin{figure}[t]
\centerline{\epsfxsize=10cm \epsfbox{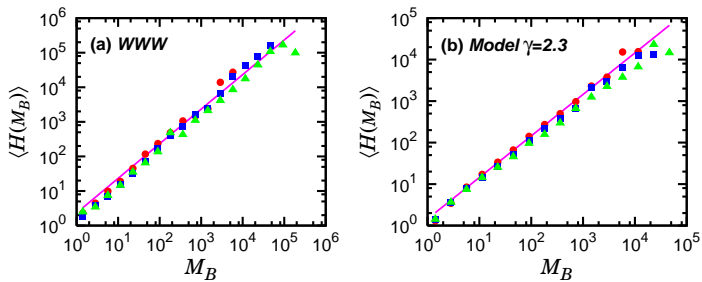}} \caption{(Color
online) Plot of the average perimeter $\langle H(M_B) \rangle$
versus box mass $M_B$ for the WWW (a) and the model network with
the parameters of $\gamma=2.3$, $p=0.5$, and $q=0.0$ (b). Data are
for box sizes $\ell_B=2$ (\textcolor{red}{$\bullet$}), $\ell_B=3$
(\textcolor{blue}{$\blacksquare$}), $\ell_B=5$
(\textcolor{green}{$\blacktriangle$}). The solid line (drawn as
reference) has a slope of 1.0 for both (a) and (b).}
\label{kprime}
\end{figure}

The power-law relation (\ref{kMBalpha}) is tested for the WWW and
the network model generated with $\gamma=2.3$ and $\langle n
\rangle=1$. It appears that $\langle H(M_B) \rangle$ depends on
the box mass $M_B$ linearly, i.e., $d_H/d_B \approx 1$,
irrespective of $\ell_B$ for the WWW [Fig.~\ref{kprime}(a)];
however, it depends on $\ell_B$ weakly for the fractal network
model. For $\ell_B=2$ and 3, $d_H/d_B \approx 1$; however, for
$\ell_B=5$, $d_H/d_B$ is likely to be marginally smaller than 1
[Fig.~\ref{kprime}(b)]. The linear behavior implies $d_H=d_B$, and
is observed in the connections between the percolation clusters
near the critical point for SF networks~\cite{havlin10}.

\section{Conclusions}

Recently, it was shown that some SF networks exhibit fractal
scaling, $N_B(\ell_B)\sim \ell_B^{-d_B}$, where $N_B(\ell_B)$ is
the number of boxes needed to tile the entire network with boxes
of size $\ell_B$. In this paper, we have introduced a modified
version of the box-covering method, which makes implementation
easy. The origin of fractal scaling is understood from the
perspective of criticality and supercriticality of the skeleton
embedded underneath each fractal SF network. By performing the
analysis of the average box mass as a function of the box size for
the box-covering and cluster-growing methods, and the mean
branching number as a function of the distance from the root, we
found that the skeleton of the WWW is a supercritical branching
tree, while the skeletons of other biological networks such as the
metabolic network of {\em E. coli} and the protein interaction
networks of {\em H. sapiens} and {\em S. cerevisiae} are critical
branching trees. Based on this observation, we constructed the
fractal network model. The box mass is heterogeneous and they
exhibit a fat-tailed behavior, $P_m(M)\sim M^{-\eta}$. We found
that the exponent $\eta$ depends on the lateral size $\ell_B$ of
the box. When $\ell_B$ is small, $\eta$ is equal to the degree
exponent $\gamma$; on the other hand, as $\ell_B$ increases,
$\eta$ approaches the exponent $\tau=\gamma/(\gamma-1)$ for the
cluster-size distribution of the branching tree; this can be
predicted from the skeleton. Finally, we studied the number of
edges that interconnect a given box and other boxes, forming the
perimeter of a box, as a function of the box mass. It appears that
the perimeter depends on the box mass linearly, and the perimeter
exponent is equal to the fractal dimension.\\

This work was supported by KRF Grant No.~R14-2002-059-010000-0 of
the ABRL program funded by the Korean government (MOEHRD) and the
Seoul R \& BD program. J.S.K.\ is supported by the Seoul Science
Fellowship. K.-I.G.\ is supported by the NSF under
the grant NSF ITR DMR-0426737.\\

\end{document}